\documentclass[useAMS,usenatbib,usegraphicx,usedcolumn]{mn2e}
\usepackage{fix2col}

\newcommand{\CIV}{\mbox{C\,{\sc iv}}}
\newcommand{\CII}{\mbox{C\,{\sc ii}}}
\newcommand{\CIII}{\mbox{C\,{\sc iii}}}
\newcommand{\SII}{\mbox{S\,{\sc ii}}}
\newcommand{\SVI}{\mbox{S\,{\sc vi}}}
\newcommand{\SIV}{\mbox{S\,{\sc iv}}}
\newcommand{\SiIII}{\mbox{Si\,{\sc iii}}}
\newcommand{\SiIV}{\mbox{Si\,{\sc iv}}}
\newcommand{\AlIII}{\mbox{Al\,{\sc iii}}}
\newcommand{\OIII}{\mbox{O\,{\sc iii}}}
\newcommand{\NV}{\mbox{N\,{\sc v}}}
\newcommand{\NIII}{\mbox{N\,{\sc iii}}}
\newcommand{\OVI}{\mbox{O\,{\sc vi}}}
\newcommand{\OI}{\mbox{O\,{\sc i}}}
\newcommand{\MgII}{\mbox{Mg\,{\sc ii}}}
\newcommand{\HI}{\mbox{H\,{\sc i}}}

\newcommand{\HeII}{\mbox{He\,{\sc ii}}}
\newcommand{\Ho}{\mbox{H$^0$}}
\newcommand{\Lya}{\mbox{Ly\,$\rm\alpha$}}
\newcommand{\Lyb}{\mbox{Ly\,$\rm\beta$}}

\newcommand{\FeII}{\mbox{Fe\,{\sc ii}}}
\newcommand{\kms}{km~s$^{-1}$}

\newcommand{\cmmt}{cm$^{-2}$}

\newcommand{\LLedd}{\mbox{$L/L_{\rm Edd}$}}
\newcommand{\Mbh}{\mbox{$M_{\rm BH}$}}
\newcommand{\auv}{\mbox{$\alpha_{\rm UV}$}}
\newcommand{\aouv}{\mbox{$\alpha_{\rm o,UV}$}}
\newcommand{\auvs}{\mbox{$\alpha_{\rm UVs}$}}
\newcommand{\auvl}{\mbox{$\alpha_{\rm UVl}$}}
\newcommand{\vshift}{\mbox{$v_{\rm shift}$}}
\newcommand{\lamrest}{\mbox{$\lambda_{\rm rest}$}}
\newcommand{\fnrepeat}[1]{$^{\ref{#1}}$}
\newcommand{\fbalq}{\mbox{$f_{\rm BALQs}$}}

\title[Average BALQ absorption properties]{The average absorption properties of broad absorption line quasars at $800<\lambda_{\rm rest}<3000$~\AA, and the underlying physical parameters}

\author[A.~Baskin, A.~Laor and F.~Hamann]
{Alexei Baskin,$^1$\thanks{E-mail: alexei@physics.technion.ac.il} 
Ari Laor$^1$ and Fred Hamann$^2$ \\
$^1$Physics Department, Technion -- Israel Institute of Technology, Haifa~32000, Israel\\
$^2$Department of Astronomy, University of Florida, Gainesville, FL 32611-2055, USA}

\begin{document}
\date{}
\pagerange{\pageref{firstpage}--\pageref{lastpage}} \pubyear{2013}
\maketitle
\label{firstpage}

\begin{abstract}
Broad absorption line quasars (BALQs) present a large diversity in their broad absorption line (BAL) profiles. To investigate what physical parameters underlie this diversity, we construct a sample of BALQs which covers $\lamrest\approx800-3000$~\AA, based on the Sloan Digital Sky Survey DR7 quasar catalogue. The average BAL properties are evaluated by taking the ratios of average BALQ spectra to the average spectra of matched samples of non-BALQs, where the matching is based on various emission properties. We find the following properties. (i) There is no detectable Lyman edge associated with the BAL absorbing gas ($\tau<0.1$). (ii) The known increase of average absorption depth with the ionization potential extends to the higher ionization \NV\ and \OVI\ BALs. We also find that the \CIV\ BAL profile is controlled by two parameters. (i) The \HeII\ emission EW, which controls the typical velocity of the \CIV\ BAL, but does not affect the absorption depth. (ii) The spectral slope in the 1700--3000~\AA\ range (\auvl), which controls the \CIV\ peak absorption depth, but does not affect the typical velocity. The \HeII\ EW and \auvl\ also control the observed fraction of quasars that are BALQs. We suggest that a lower \HeII\ EW may indicate a weaker ionizing continuum, which allows the outflow to reach higher velocities before being overionized, possibly without a need to invoke a shielding gas. A redder continuum may indicate a more inclined system, and a larger covering factor and larger column of the outflow along the line of sight. 
\end{abstract}

\section{Introduction}\label{sec:intro}
Broad absorption line quasars (BALQs) are a subtype of active galactic nuclei (AGN), defined by the presence of broad and blue-shifted absorption features (e.g., \citealt*{weymannetal81}). The intrinsic fraction of BALQs from the total quasar population is estimated to be $\sim15-20$ per cent (\citealt{hewfol03, reichardetal03, kniggeetal08, gibsonetal09}. \citealt{allenetal11} claim it can be as high as $\sim40$ per cent). While there are several differences in emission properties between BALQs and non-BALQs, BALQs appear to be drawn from the non-BALQ population \citep*{weymannetal91, hamannetal93, reichardetal03}. Broad absorption lines (BALs), and the \CIV\ BAL in particular, span a large range in depth, width and overall velocity shift (\vshift) between different objects. For most BALQs, absorption is observed only in high-ionization lines (these objects are termed `HiBALQs'). For a smaller fraction of BALQs, absorption is also observed in low-ionization lines, e.g.\ \MgII\ (`LoBALQs'). The predominant unifying model states that the difference between the AGN subtypes is our viewing angle towards the quasar central regions (e.g., \citealt{elvis00}). However, an alternative scenario suggests that LoBALQs are an evolutionary stage of AGN, in which the nucleus expels a surrounding dusty `cocoon' (\citealt*{voitetal93, urrutiaetal09, farrahetal10, farrahetal12, glikmanetal12}; cf.\ \citealt{lazarovaetal12}).

What are the average BALQ spectral properties shortward of \Lya, and near the Lyman limit?  Most of BALQ studies analyse spectra only down to \lamrest\ of \SiIV\ (e.g., \citealt{gibsonetal09, allenetal11}) or \NV\ (e.g., \citealt{weymannetal91}). In this study, we utilize the Sloan Digital Sky Survey (SDSS; \citealt{york00}) DR7 quasar catalogue \citep{schneideretal10, shenetal11}, and investigate high-$z$ $(>3)$ object spectra, which cover \lamrest\ from \CIV\ down to $\sim800$~\AA. The spectra of $z>3$ quasars are heavily absorbed by the intervening Lyman forest at $\lamrest< 1216$~\AA. We overcome this foreground absorption and derive the BALQ intrinsic absorption by taking the ratio of the average spectrum of BALQs and the average spectrum of a matched control sample of non-BALQs. The ratio spectrum allows us to place a limit on the average Lyman edge depth associated with the BAL systems, and thus a limit on the covering factor (CF) of low ionization BAL system. We also extend earlier studies on the average absorption strength of \CII, \SiIV\ and \CIV\ to the higher ionization \NV\ and \OVI\ BALs.

The ratio spectrum is meaningful only if the average emission properties of BALQs and non-BALQs are indeed the same. Various studies find systematic differences between the intrinsic emission properties of BALQs and non-BALQs. There are reports that BALQs are located on the high-$L$ and high-\LLedd\ end of the \citet{borgre92} eigenvector 1 \citep{bor02}. BALQs are observed to have larger blueshifts of \CIV\ emission than non-BALQs \citep{richardsetal02, reichardetal03}, and LoBALQs to have the highest blueshifts \citep{richardsetal02}. BALQs are found to be redder than non-BALQs \citep{reichardetal03, maddoxetal08, gibsonetal09, allenetal11}, and LoBALQs to be redder than HiBALQs \citep{weymannetal91, sprfol92, bormey92, reichardetal03, gibsonetal09}. The reddening is interpreted by some authors as a result of BALQs being viewed preferentially closer to edge on (e.g., \citealt{ogleetal99}), or as an imprint of a dusty `cocoon' (see references above). \citet{trumpetal06} find the emission lines to be broader for BALQs than for non-BALQs. BALQs are reported to have lower \CIV\ equivalent width (EW) than non-BALQs, X-ray weaker BALQs are observed to have stronger BALs with larger terminal velocities, and the measured velocities are larger for higher-$L_{\rm UV}$ BALQs \citep*{brandtetal00, laobra02, gibsonetal09}. LoBALQs show the strongest and broadest high-ionization absorption lines \citep{allenetal11}.

Thus, to make a more accurate ratio spectrum one needs to use samples of BALQs and non-BALQs with a similar distribution of intrinsic emission properties. One can then take another step, and explore whether the derived absorption properties, in particular the average \CIV\ absorption profile, depend on the intrinsic emission properties, such as \LLedd. For this purpose, we expand our study to $z\sim 1.5$ quasars, where one can observe the 1400--3000\AA\ range, and derive emission parameters, such as \LLedd, based on the \MgII\ $\lambda$2798 broad emission line. Despite numerous studies, the intrinsic properties which underlie the diversity of BALs remain elusive. This study allows us to address what causes the large diversity of the observed BAL properties 

We explore in this study the dependence of the \CIV\ BAL properties on the \HeII\ EW and on the UV slope (\auv). This exploration is motivated by the following. The \HeII\ EW is a measure of the strength of the extreme UV (EUV) continuum above 54~eV, compared to the near UV continuum, and there are reports that BALQs have on average lower \HeII\ EW than non-BALQs \citep{richardsetal02, reichardetal03}. A Broad Line Region (BLR) wind component is reported to be affected by the ionizing continuum hardness (\citealt{leimoo04}, based on \HeII\ EW; \citealt{kruczeketal11}, based on $\alpha_{\rm ox}$), and the BLR wind component is suggested by \citet{richards12} to be possibly relevant to the BALQ phenomenon. In addition, \auv\ is correlated with reddening (e.g., \citealt{bl05,sl12}), and since reddening is more common in BALQs (see above), this implies a possible relationship between \auv\ and BALQ properties. 

The paper is structured as follows. The data analysis method is described in Section~\ref{sec:methods}. In Section ~\ref{sec:mean_prop} we analyse composite spectra of the BALQ and non-BALQ samples, find a trend between the ionization potential and the average BAL depth, and constrain the average \Ho\ absorber properties. In Section~\ref{sec:what_is_driver} we investigate what parameters span the \CIV\ BAL properties. A physical interpretation to our findings is proposed in Section~\ref{sec:phys}. In Section~\ref{sec:dust} we examine which dust extinction laws can explain the BALQ reddening relative to non-BALQs. Our conclusions are summarized in Section~\ref{sec:conclude}.

\section{The Data Analysis}\label{sec:methods}
The data set is drawn from the SDSS DR7. The object BALQ classification is adopted from the \citet{shenetal11} quasar catalogue.\footnote{The updated \citet{shenetal11} quasar catalogue, available at http://das.sdss.org/va/qso\_properties\_dr7/dr7.htm, includes the improved $z$ from \citet{hewwil10}.} \citet{shenetal11} use the \citet{gibsonetal09} BALQ classification for objects that are included in the SDSS DR5, and classify the remaining objects based on a visual inspection of the \CIV\ region. Note that \citet{gibsonetal09} use a modified version of `balnicity index' (BI) of \citet{weymannetal91} to detect BALQs, which they term BI$_0$. They integrate the continuum-normalized spectral flux starting from a blueshift of 0, rather than $-3000$~\kms\ used in the traditional BI. We include in the data set only objects with $\mbox{S/N}\geq3$ in the SDSS \textit{i}-filter, to avoid unusually low-S/N spectra (the S/N criterion excludes $\sim5-10$ per cent of the objects; see below). We divide the data set into two subsets that cover different \lamrest\ ranges as described below.
\begin{enumerate}
\item The $800\leq\lambda_{\rm rest}\leq1750$~\AA\ range i.e., $3.75\leq z\leq4.25$ for the SDSS (hereafter, the `high-$z$' sample). The lower limit on $z$ is set to allow a detection of Lyman limit absorption intrinsic to the BALQs, and the upper limit is set to detect the continuum redward of the \CIV\ emission complex. The DR7 covers this $z$ range for 228 BALQs and 1320 non-BALQs. The $\mbox{S/N}\geq3$ criterion leads to 200 BALQs and 1142 non-BALQs.

\item The $1400\leq\lambda_{\rm rest}\leq3000$~\AA\ range i.e., $1.75\leq z\leq2.05$ for the SDSS (hereafter, the `low-$z$' sample). The lower and upper limits are placed to cover the \CIV\ BAL and the \MgII\ emission line, respectively. The DR7 contains 1691 BALQs and 13,388 non-BALQs in this range. The S/N criterion excludes 39 BALQs and 739 non-BALQs. Since LoBALQs are a distinct subtype of BALQs, with redder spectra than the more common HiBALQs \citep{weymannetal91, sprfol92, bormey92, reichardetal03, gibsonetal09}, we exclude from the low-$z$ BALQ sample 56 objects with a detected \MgII\ BAL \citep{shenetal11}, and construct a sample of low-$z$ HiBALQs only. We do not construct a similar high-$z$ HiBALQ sample, as it is not clear which low-ionization absorption line at $\lamrest\leq1750$~\AA\ matches the \MgII\ absorption line. Note that \citet{shenetal11} do not conduct a systematic search for \MgII\ BALQs in the post-DR5 quasar sample, and report only serendipitously found \MgII\ BALQs for this sample. Thus, the exclusion of LoBALQs from the low-$z$ BALQ sample might be incomplete. \citet{trumpetal06} report a \MgII\ BALQ fraction of 1.3 per cent of quasars for DR3 (i.e.\ 164 objects in our sample), where BALs are detected using the `absorption index' $\textrm{AI}>0$ criterion, while \citet{allenetal11}, who use the $\textrm{BI}>0$ criterion, find a smaller fraction of 0.3 per cent for DR6 (38 objects). The different fractions result from the different definitions of AI and BI.\footnote{$\textrm{AI}>0$ ($\textrm{BI}>0$) requires at least one continuous absorption trough with a minimal absorption depth of 0.1 and a minimal width of 1000~\kms\ (2000~\kms) in the $-29,000<\vshift<0$~\kms\ ($-25,000<\vshift<-3000$~\kms) range.} Thus, our low-$z$ HiBALQ sample is likely contaminated by $\sim 100$ LoBALQs which pass the $\textrm{AI}>0$ criterion, but none of these pass the $\textrm{BI}>0$ criterion. The final low-$z$ sample is comprised of 1596 HiBALQs and 12,649 non-BALQs.
\end{enumerate}

Since the spectra of high-$z$ objects have a relatively low S/N (a median value of $\sim7$ per resolution element at continuum regions unaffected by absorption and strong emission)\footnote{The S/N of an object is estimated by calculating the ratio between the mean and standard deviation of $f_\lambda$ in the $\lambda_{\rm rest}=1700-1720$~\AA\ window.}, the spectra are smoothed by a 22 pixel-wide moving average filter, which allows to achieve a median S/N of $\sim22$ per resolution element [the filter width is equivalent to $\sim9$ resolution elements i.e., $\sim1350$~\kms\ \citep{york00}]. The spectra of low-$z$ objects are smoothed by the same filter, yielding a median S/N of $\sim66$ per resolution element. Since we are dealing with BALs (i.e., line width $>2000$~\kms), the absorption profiles explored in this study remain well resolved. All spectra are normalized by the mean flux density in the $\lamrest=1700-1720$~\AA\ range.\footnote{It should be noted that the wavelength used for the normalization is not significant, as verified by repeating the main analysis using the $\lamrest=1275-1285$~\AA\ range instead.}

Median spectra are calculated by utilizing a modified median method. A simple mean is not adopted as the representative average spectrum, since it can be affected by outliers. Outliers have a negligible effect on the median, but the median yields a relatively `noisier' composite compared to the mean. We utilize a modified median method, which is a hybrid between the standard median and the mean, and which mitigates the shortcomings of the two methods. First, the object spectra are sorted based on the normalized $f_{\lambda}(\lamrest)$. Then, 10 per cent of the objects above and 10 per cent of the objects below the median are marked. Finally, the mean $f_{\lambda}$ of the marked objects is adopted as the composite $f_{\lambda}$ at a given \lamrest\ (a similar method is used to calculate the Libor interest rate in Economy). 

The intrinsic BALQ absorption is evaluated by utilizing the non-BALQ sample. The non-BALQ composite is assumed to represent the intrinsically unabsorbed emission of the BALQs (the construction of composites matched in emission properties is described below). The ratio between the BALQ and non-BALQ composites (hereafter the $R$ spectrum) is interpreted as the intrinsic BALQ absorption spectrum. Since the ratio does not remove an overall spectral energy distribution (SED) differences between BALQs and non-BALQs, we fit a straight-line continuum across the profile from $\vshift=0$ to $-30,000$~\kms. A residual \SiIV\ emission, if present, has little effect on the \CIV\ measurements. The caveat of our approach is that if there are significant intrinsic differences in line emission within the absorbed region for BALQs, then the derived absorption profile will be biased.

The individual absorption profiles are often characterized by a relatively narrow absorption peak. This characteristic profile is smeared out when forming the composite, due to the distribution of peak absorption velocities. In order to investigate the median `BAL rest-frame' absorption profile near the absorption peak,  we also form a composite based on the aligned \CIV\ peak-absorption. This composite, calculated for the high-$z$ BALQs, allows to detect relatively narrow ($\textrm{FWHM}\ga1500$~\kms) and weak absorption lines which are otherwise heavily blended in the crowded FUV region, and also provides some hint on the optical depth of resolved multiplet absorption. The BALQ spectra are aligned by shifting each spectrum in velocity, so that the maximum absorption of \CIV\ falls at the same velocity for all spectra. Then, a composite of the aligned spectra is calculated. The 22-pixel spectral smoothing by a relatively broad filter, smooths out any narrow features ($\la500$~\kms) superimposed on the \CIV\ BAL, and reduces their effect on the alignment procedure. When the matched non-BALQ sample is constructed, a similar distribution of velocity shifts is used. 

We study the dependence of the BALQ absorption properties on the following parameters. 
\begin{enumerate}
\item The \HeII\ EW, which may serve as a measure of the EUV ionizing SED hardness. The \HeII\ EW is measured by integrating the normalized $f_\lambda$ in the $\lambda_{\rm rest}=1620-1650$~\AA\ range, assuming $f_\lambda^{\rm cont}=1$ i.e., a constant $f_\lambda^{\rm cont}$ between the normalization window (1700--1720~\AA) and $\sim1620$~\AA. The adopted \lamrest\ range minimizes contributions from \CIV\ and \OIII] $\lambda\lambda$1661, 1666 to the evaluated \HeII\ EW. Using a constant $f_\lambda^{\rm cont}$ corresponds to a spectral slope of $\alpha=-2$ ($f_\nu\propto\nu^{\alpha}$), which serves as a rough approximation for the local slope (Section~\ref{sec:trends_auv}). A significantly more accurate derivation of the local $f_\lambda^{\rm cont}$ is hindered by the \CIV\ BAL for BALQs. The derived \HeII\ EW is slightly overestimated in objects where the continuum slope is bluer, but these objects tend to have a higher \HeII\ EW, where the exact continuum placement is less significant. The constant $f_\lambda^{\rm cont}$ is a good approximation in the low \HeII\ EW objects. In the reddest objects, the continuum at $\lamrest<1700$~\AA\ drops below the adopted constant $f_\lambda^{\rm cont}$, leading to slightly negative \HeII\ EW (see Section~\ref{sec:trends_both}). The small errors in the absolute values of the \HeII\ EW are significantly smaller than the trends with the \HeII\ EW explored below.

\item The UV spectral slope $\alpha_{\rm UV}$, which may serve as a measure of dust absorption. For the high-$z$ sample, the slope is evaluated between the 1275--1285 and 1700--1720~\AA\ windows, and is denoted as \auvs. The latter window is redward of \OIII] $\lambda$1665 emission and the \CIV\ emission `shelf', and blueward of \NIII] $\lambda$1750. The former window is between the red wing of \Lya+\NV\ emission complex and \OI~$\lambda$1303. The window between \OI~$\lambda$1303 and \CII\ $\lambda$1335 is not utilized, because it is affected by \CII\ and \SiIV\ absorption in BALQs. For the low-$z$ sample, the slope is evaluated between the 1700--1720 and 2990--3010~\AA\ windows, and is denoted as \auvl. The latter window is redward of \MgII\ $\lambda$2798.
\end{enumerate}
We also explore the dependence of the BALQ absorption, for the low-$z$ sample, on the following additional parameters.
\begin{enumerate}
\item $L(3000\mbox{\AA})$, taken from \citet{shenetal11}.
\item \MgII\ FWHM, taken from \citet{shenetal11}.
\item \Mbh, estimated using parameters (i) and (ii), and the \citet{vesosm09} prescription (values listed in \citealt{shenetal11}).
\item \LLedd,  evaluated as $\log\LLedd=\log L(3000\mbox{\AA})-\log\Mbh-37.4$, where we adopt a bolometric correction factor of 5.15 \citep{shenetal08}.
\end{enumerate}
We do not use the \CIV\ emission line to estimate \Mbh\ given the significant uncertainties associated with this line (e.g.\ \citealt{bl05}, and citations thereafter). The objects are binned for each parameter, so that each BALQ bin contains the same number of objects. The corresponding non-BALQ bins span the same parameter range, but the distribution of objects within each bin may be different. A median composite spectrum is calculated for each bin.

\section{The average BALQ properties}\label{sec:mean_prop}
\begin{figure*}
\includegraphics[width=174mm]{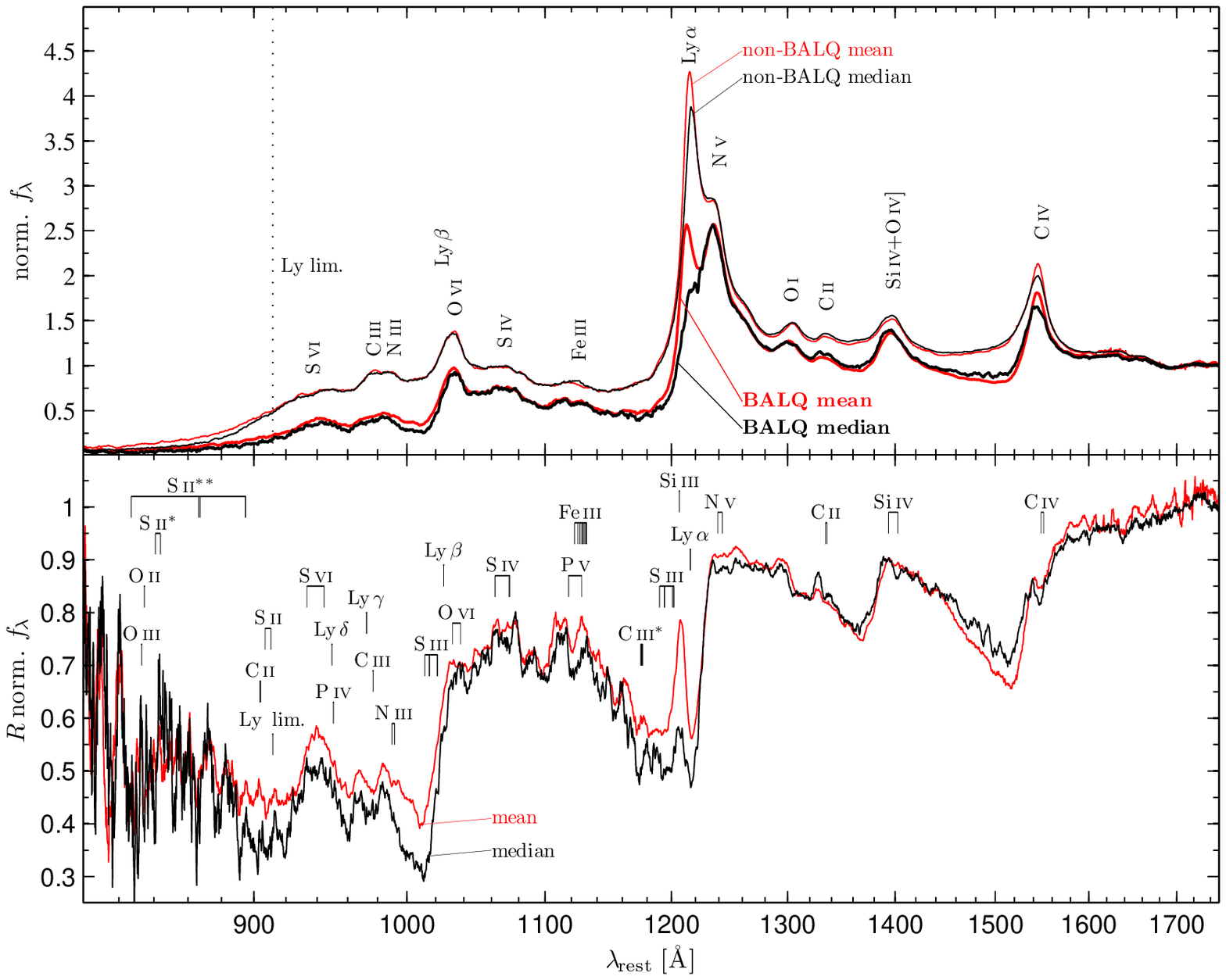}
\caption{Comparison between the BALQ and non-BALQ composite spectra. Top panel:  the median normalized flux of the BALQ and non-BALQ samples (thick and thin black solid line, respectively). The mean normalized flux is also presented for comparison (red lines). The composites can differ at strong absorption and emission features (e.g., \Lya), where the mean is more affected by outliers. The Lyman limit (vertical dotted line) and prominent emission lines are indicated. Bottom panel: the ratio between the BALQ and non-BALQ composites. The laboratory wavelength location of prominent absorption lines is indicated longward of the Lyman limit. We also indicate for reference the location of possible lines below the Lyman limit. The indicated \SII$^*$ and \SII$^{**}$ lines have excitation energy of 1.8 and 3.1~eV, respectively. There is no significant absorption edge detected at the Lyman limit. Note also the declining continuum ratio to the blue, and the increasing absorption features depths.}\label{fig:all}
\end{figure*}

Figure~\ref{fig:all} presents a comparison between the high-$z$ BALQ and non-BALQ composites, aligned by their tabulated $z$. We present both the median and the mean composites, to demonstrate the level of systematics produced by the averaging procedure, which is larger than the purely statistical noise. The two composites are similar overall, except at wavelength regions of strong absorption and emission features (e.g., \Lya), where the mean composite is more affected by outliers. Bottom panel presents the $R$ composite. The median and mean $R$ composites are similar. There are small differences in absorption amplitude, which are likely selection effects.\footnote{The reversal of the mean versus median $R$ spectra at the \CIV\ and \OVI\ absorption troughs is not a significant result. It reflects the fact that the residual flux at the \OVI\ trough is closer to zero than for \CIV.} We indicate the laboratory wavelength location of absorption lines of various ionized species. There are many absorption lines in the $\lambda_{\rm rest}=800-900$~\AA\ range, and the indicated absorption line locations are for illustration only. The steep decline of flux blueward of the Lyman edge for both BALQ and non-BALQ composites (top panel) is caused by intervening Lyman $\alpha$ systems. Observations of low-$z$ AGN, which are unaffected by intervening Lyman systems, do not show this steep decline \citep{zhengetal97, telferetal02, scottetal04, shulletal12}. 

The three main features which can be seen in the $R$ composites are: (i) there is no detectable Lyman limit absorption; (ii) the high ionization lines observed shortward of \Lya, in particular \NV\ and \OVI, show stronger absorption than observed longward of \Lya; (iii) the overall continuum ratio is red. These results are further discussed below.

\begin{figure}
\includegraphics[width=84mm]{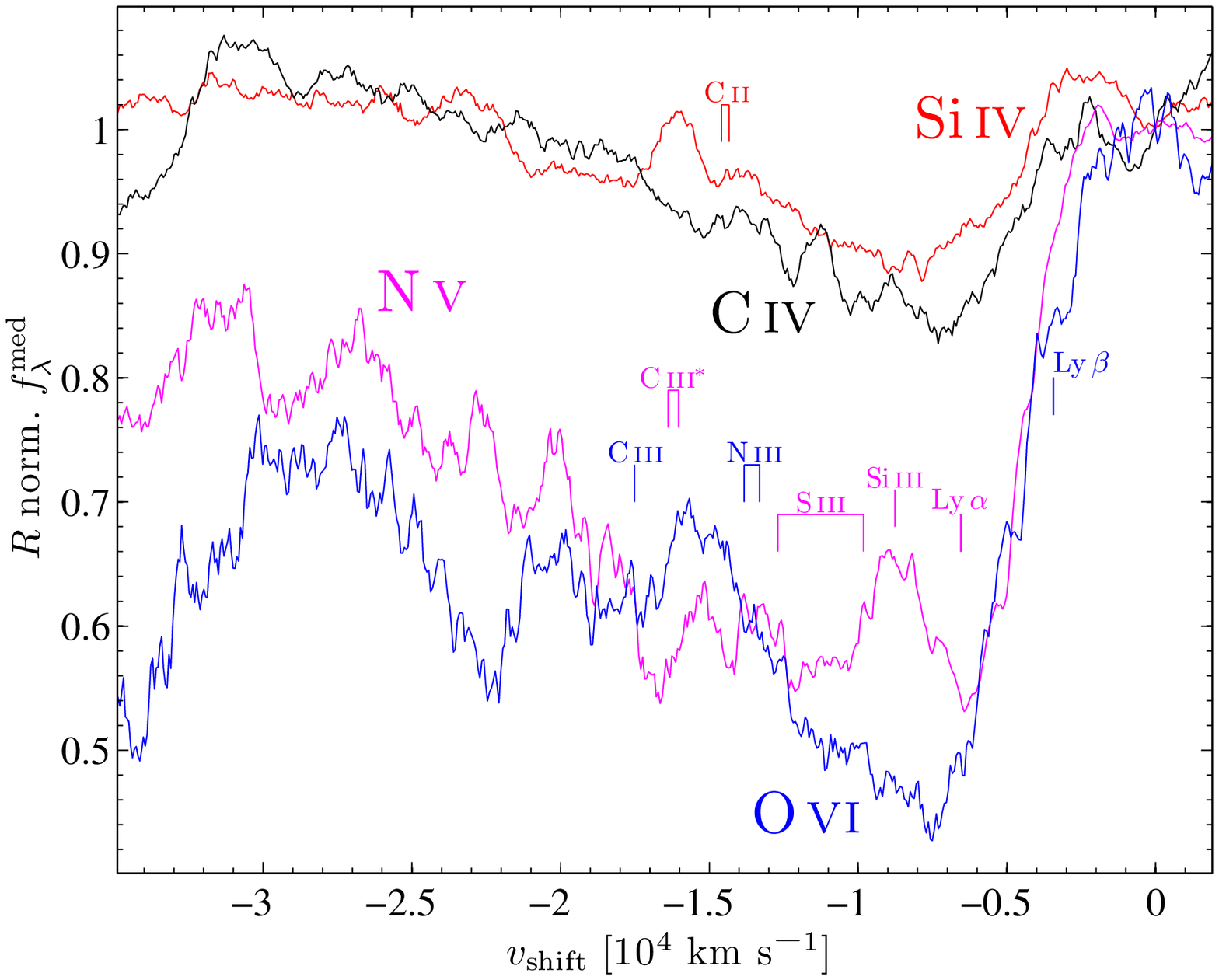}
\caption{The absorption profiles of the prominent absorption lines \SIV, \CIV, \NV\ and \OVI. The velocity scale is set by the longer wavelength of each doublet. The profiles are normalized to 1 at $v_{\rm shift}=0$~\kms. Laboratory wavelength location of various other possible absorption lines is indicated by vertical tick marks (with the same colour coding). The absorption depth increases with the ionization potential of the absorbing ion.}\label{fig:zoom_on_abs}
\end{figure}

Figure~\ref{fig:zoom_on_abs} compares the absorption profiles of \SiIV, \CIV, \NV\ and \OVI.  The higher the ionization potential, the stronger is the BAL trough. The \SiIV\ absorption line, produced by photons above 33.5~eV and destroyed by photons above 45.1~eV, has the shallowest trough. It is followed in trough depth and absorption EW by \CIV, \NV\ and \OVI\ (47.9--64.5, 77.5--97.9 and 113.9--138.1~eV, respectively). The absorption EW (transmission at maximal absorption) is 5.8 (0.87), 9.3 (0.83), 24.6 (0.56) and 25.2~\AA\ (0.46) for \SiIV, \CIV, \NV\ and \OVI, respectively. The \NV\ absorption may be affected by \Lya\ at $\vshift=0$. Note that \CII\ $\lambda$1335 falls on the same trend, as its absorption is weaker ($\textrm{EW}\la1$~\AA) than \SiIV\  (Fig.~\ref{fig:all}, bottom panel). 

Our findings are consistent with the \citet{allenetal11} report that on average \SiIV\ reaches smaller values of peak absorption than \CIV. \citet{gibsonetal09} and \citet{allenetal11} find that the \CIV\ BI distribution has a marginally larger tendency towards higher values than \SiIV. The difference in absorption depth likely implies difference in CF. 

\begin{figure*}
\includegraphics[width=174mm]{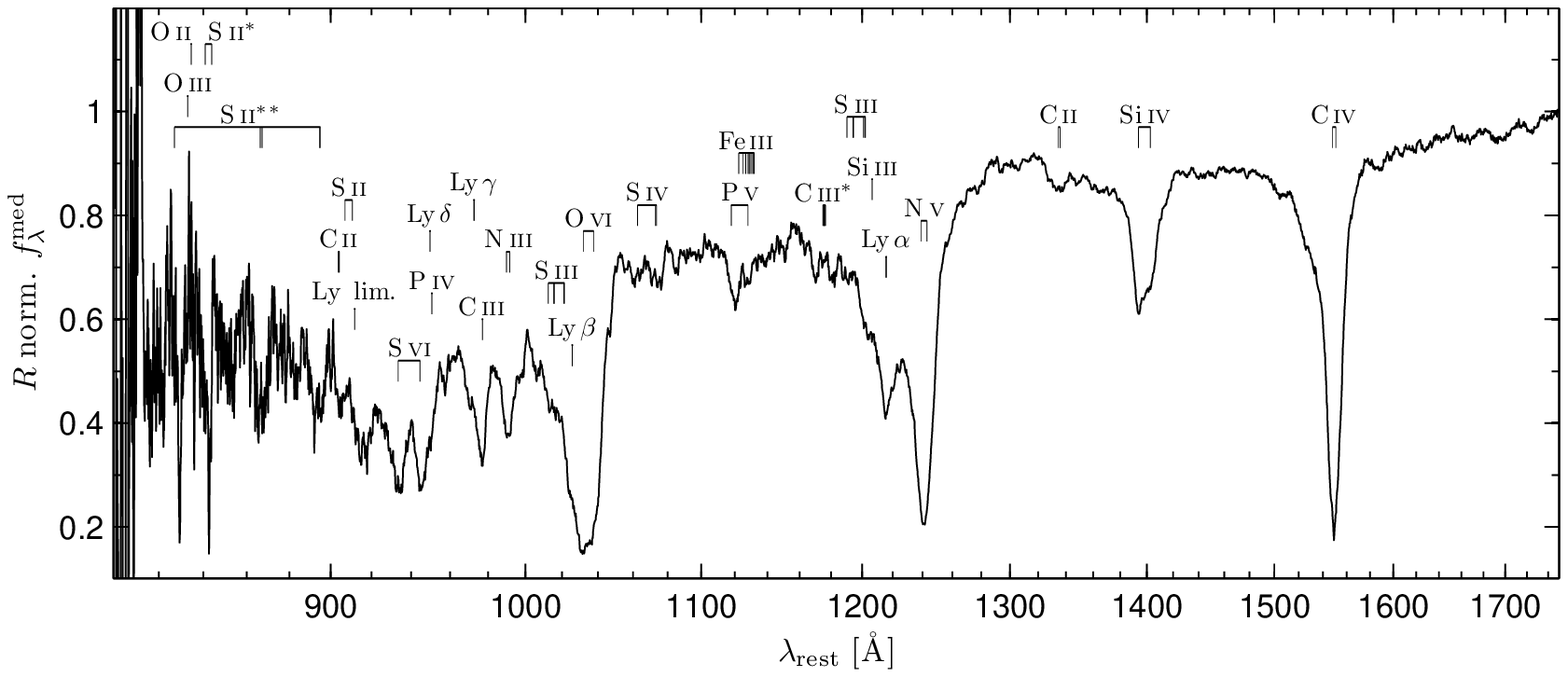}
\caption{The absorption profiles near the absorption peak. The spectra are aligned so that the \CIV\ absorption dip falls at 1549.48~\AA\ (i.e., a 1:1 contribution to absorption from the two components of \CIV\ $\lambda\lambda$1548.20, 1550.77). Note that both the low- and the high-ionization BALs are at the same $v_{\rm shift}$ as \CIV. This allows to separate heavily blended features, such as \Lya\ and \NV, and to detect weak and relatively narrow features, which are otherwise heavily blended, such as \SVI\ (Fig.~\ref{fig:all}).}\label{fig:R_aligned}
\end{figure*}

Figures~\ref{fig:R_aligned} and \ref{fig:abs_prof_alig} present the \CIV\ peak-absorption aligned $R$ spectrum and absorption line profiles, respectively. Line identifications are indicated in Fig.~\ref{fig:R_aligned} at the line laboratory wavelength. The peak-absorption alignment method aligns all the BALs, including low-ionization BALs (e.g., \CII\ $\lambda$1335). The line profiles of the BALs are examined further in Fig.~\ref{fig:abs_prof_alig}, where the lines are grouped vertically based on the ionization potential. The absorption profiles are dominated by the relatively narrow absorption component ($\textrm{FWHM}\approx3000$~\kms). Fig.~\ref{fig:abs_prof_alig} indicates that for several lines the absorption is saturated, as indicated by the similar absorption depth of doublet components with different oscillator strengths. The absorption profile in these cases is set by a velocity dependent CF (e.g., \SVI\ $\lambda\lambda$933, 945). Other lines are not saturated (e.g., \SiIV\ $\lambda\lambda$1397, 1403), as can be indicated by more quantitative analysis (e.g. \citealt{dunnetal12}, Capellupo et al.\ in prep.\ and Hamann et al.\ in prep.).

\subsection{Constraints on the BAL H$^0$ Column}\label{sec:const_ho}
\begin{figure*}
\includegraphics[width=174mm]{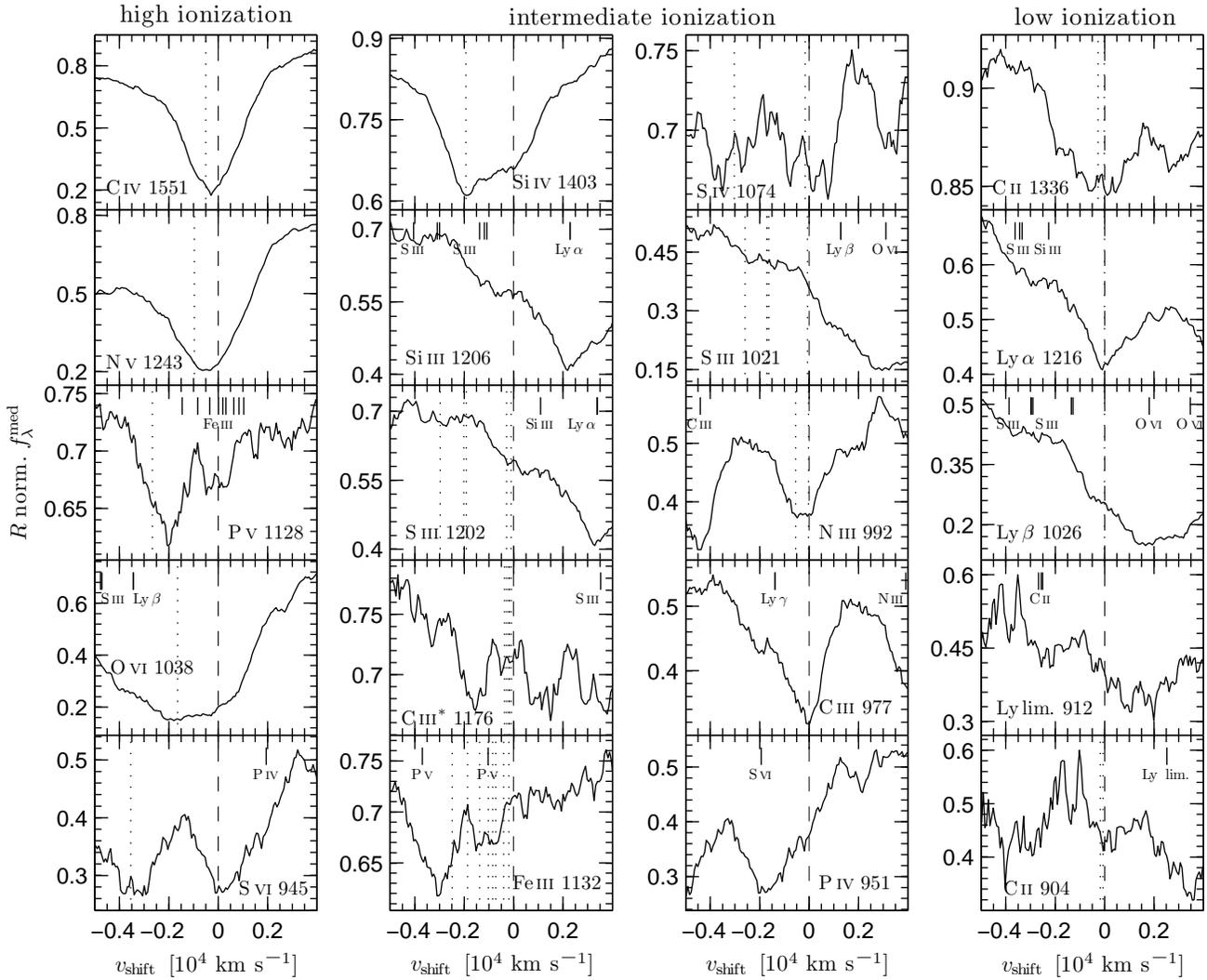}
\caption{Line profiles of the aligned composite spectrum (see Fig.~\ref{fig:R_aligned}). The lines are grouped according to the ionization potential, and are arranged from the longer to the shorter wavelength at each group. The velocity scale is calculated relative to the redder line of each multiplet, which is indicated at each panel. A zero velocity shift and $v_{\rm shift}$ of other absorption lines of each multiplet are marked (dashed and dotted line, respectively). The location of other absorption lines is indicated by vertical tick marks at the top of each panel. Note that some of the doublets appear optically thick (e.g., \SVI~945), while others appear unsaturated (e.g., \SiIV~1403).}\label{fig:abs_prof_alig}
\end{figure*}

Figure~\ref{fig:abs_prof_alig} can be utilized to constrain the average BAL \Ho\ column $N(\Ho)$, based on the \Lya\ absorption line. We begin by deriving $N(\Ho)$ assuming a uniform foreground screen i.e., ${\rm CF}=1$. The \Lya\ absorption line resides within the \NV\ absorption trough. We estimate the \Lya\ absorption depth by interpolating linearly between the two `shoulders' at $\vshift=\pm2000$~\kms\ (Fig.~\ref{fig:abs_prof_alig}), which yields a continuum $f_\lambda\approx0.55$ at the line centre. The measured $f_\lambda\approx 0.42$ at the peak absorption implies a \Lya\ peak absorption of $\exp(-\tau_0)=0.42/0.55\approx 0.76$. Assuming a Maxwellian absorption profile with a velocity dispersion $b$ parameter of 1000~\kms, yields $N(\Ho)\simeq 1\times10^{15}$~\cmmt\ (e.g., \citealt{draine11}, eq.~C.1). This $N(\Ho)$ implies $\tau\simeq 0.01$ at the Lyman edge (e.g., \citealt{osterbrock89}, eq.~2.4), which is consistent with a non-detection of the edge in the absorption-peak aligned $R$ spectrum (Fig.~\ref{fig:abs_prof_alig}, Ly lim.\ panel). However, the \Lya\ line and Lyman edge can be highly saturated, with $N(\Ho)$ well above the derived value. In that case, the \Lya\ absorption profile reflects the CF of the saturated cold absorber, which is at least $1-0.76=0.24$. However, the observed Lyman edge region shows no detectable edge, i.e. $\tau\la 0.1$, and implies $N(\Ho)< 1\times10^{17}$~\cmmt\ if $\textrm{CF}=0.24$. In addition, the \Lyb\ shows only weak absorption, well below 0.24, if we adopt the \OVI\ shoulder as the local continuum, which also suggests a non saturated \Lya\ absorption.

Another constraint on the CF of any nearly neutral saturated absorber can be derived from the \CII\ $\lambda$1335 absorption line. The $R$ spectrum in the \CII\ region shows $f_\lambda\approx0.84$ at the line centre, compared to an estimated local continuum of 0.89 (Fig.~\ref{fig:abs_prof_alig}, \CII\ 1336 panel), which implies a ${\rm CF}\la 0.06$, consistent with the non detection of the expected Lyman edge absorption.

We note in passing that photoionization models indicate that the \Lya\ absorption should be at least as strong as \SiIV\ for solar metallicity, regardless of the ionization parameter \citep{bl12}. This is consistent with our observations, as the \SiIV\ absorption goes down from $f_\lambda\approx0.9$ to 0.62 that indicates peak absorption of $\sim0.3$, which is similar to the \Lya\ peak absorption. 

An obvious general caveat is that the above analysis is applied to the median absorption profile, and may not be valid to individual objects, such as LoBALQs, which are not excluded from the high-$z$ BALQ sample.

\section{Which parameters control the absorption profile?}\label{sec:what_is_driver}
In this section we use composite spectra to examine the relationship of \CIV\ BAL properties to \HeII\ EW, \auv, and other measured parameters.

\subsection{\CIV\ BAL trends with the \HeII\ EW}\label{sec:trends_EW}
\begin{figure*}
\includegraphics[width=174mm]{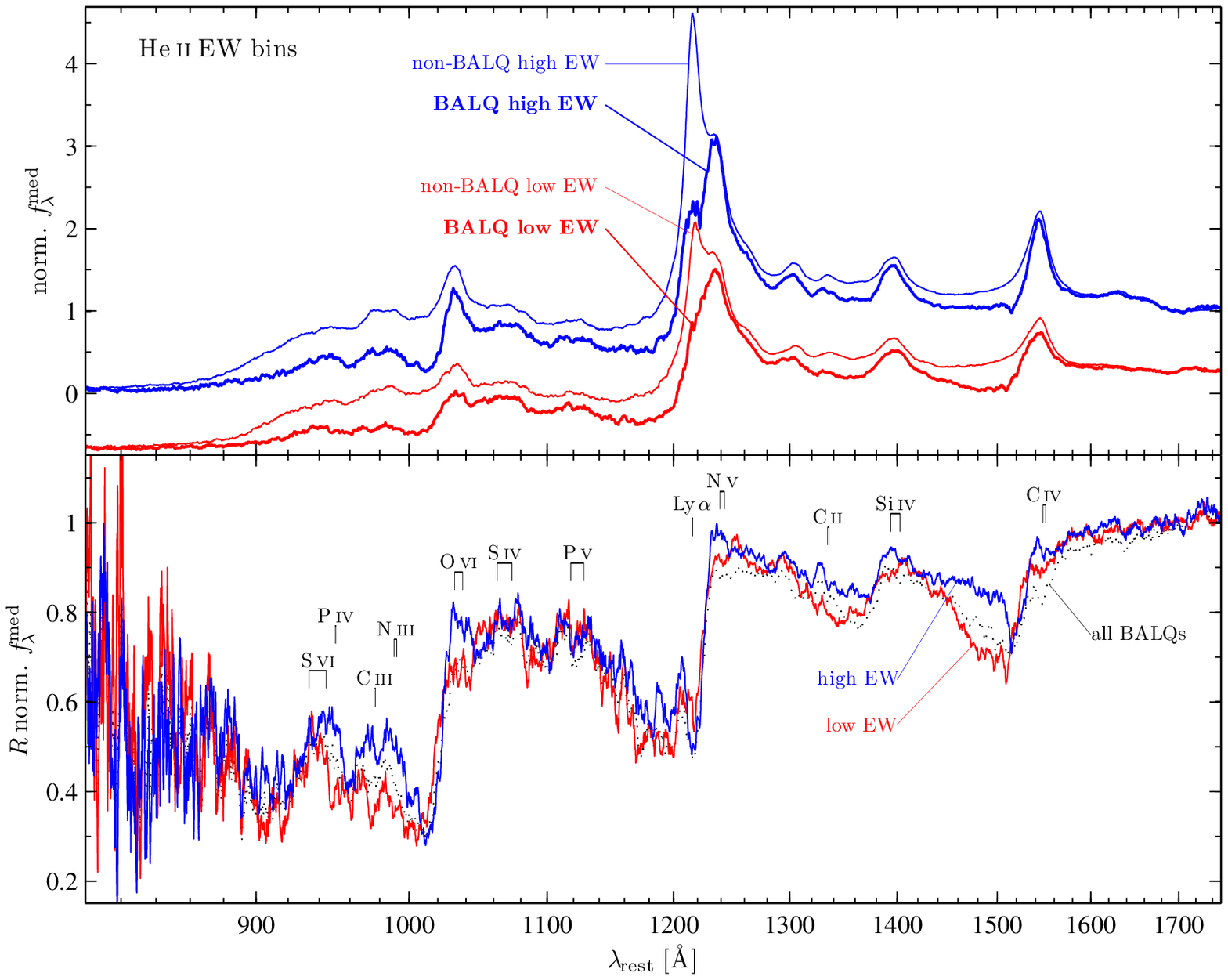}
\caption{The dependence of the intrinsic absorption on the \HeII\ EW for the high-$z$ sample. Top panel: the composite spectra of high and low \HeII\ EW BALQs and non-BALQs. The low \HeII\ EW spectra are shifted down by 0.7 for presentation purposes. Bottom panel: the ratio between the corresponding BALQ and non-BALQ bins. The ratio of the complete samples is also presented (same as in Fig.~\ref{fig:all}), which shows an apparent red side absorption in \CIV\ that results from the different median \CIV\  emission profiles in BALQs and non-BALQs. This effect is partly compensated by using the two subsamples, which have better matched \CIV\ emission strengths. Note the significantly stronger \CIV\ absorption for the lower \HeII\ EW composite. A similar effect in seen in the \SIV\ BAL, some effect may be seen in the \OVI\ BAL, and only a marginal effect is seen in \NV, possibly due to blending with the low ionization \Lya\ BAL.}\label{fig:EW_subs_lt1700}
\end{figure*}

Figure~\ref{fig:EW_subs_lt1700} presents the high-$z$  BALQ and non-BALQ samples separated into two bins based on their \HeII\ EW (as further described below). The upper panel presents the composite spectra of the two bins for BALQs and for non-BALQs, and the lower panel presents the $R$ spectra for the two bins. The low \HeII\ EW bin displays a somewhat deeper and significantly broader \CIV\ BAL, with peak absorption shifted to larger \vshift. This trend is also observed for \SiIV\ and possibly for \OVI, although a clear detection of the trend for the latter is hindered by the low S/N and the blending of a number of adjacent absorption lines. The trend does not appear significant for \NV, probably due to contamination by the adjacent low ionization line \Lya, which is prominent relative to other low ionization lines (e.g., Fig.~\ref{fig:R_aligned}). The weaker trend between the \HeII\ EW and the \NV\ and \OVI\ BALs is further discussed in Section~\ref{sec:phys_EW}. The drop in the $R$ spectrum at the red wing of the \CIV\ emission line reflects a mismatch in the \CIV\ emission EW between BALQs and non-BALQs for the same bin (see Fig.~\ref{fig:EW_subs_lt1700}, top panel), and is most likely not an absorption effect. Table~\ref{tab:HeII_highz} summarizes various properties of the two \HeII\ EW bins, and lists the first three moments of the \CIV\ absorption profile (EW, mean \vshift\ and dispersion $\sigma$).

\begin{table*}
\begin{minipage}{110mm}
\caption{The median properties of the \HeII\ EW binned high-$z$ objects.}\label{tab:HeII_highz}
\begin{tabular}{@{}{l}*{7}{c}{c}@{}}
\hline
Class & Bin & $N_{\rm obj}$ & \HeII &\auvs\fnrepeat{fn1:auv} & \multicolumn{3}{c}{\CIV\ BAL\fnrepeat{fn1:EW}} & \fbalq\fnrepeat{fn1:balq}\\
&  &  & EW &&  EW & $v_{\rm shift}^{\rm mean}$ & $\sigma$ \\
& &  & [\AA] && [\AA] & [\kms] & [\kms] & [per cent]\\
\hline
BALQs & high & 100 & 5.4 & $-1.04$& 12.0 &$-$13000 & 6800 & $11\pm1$\\
 & low & 100 & 0.4 & $-1.76$ & 14.8 & $-$12500 & 4700 & $22\pm2$\\
\hline
non- & high & 778 & 6.1 & $-0.71$\\
BALQs & low & 353& 0.7 & $-1.36$\\
\hline
\end{tabular}
\footnotetext[1]{Measured between 1280 and 1710~\AA\ ($f_\nu\propto\nu^{\alpha}$).\label{fn1:auv}}
\footnotetext[2]{Measured between $v_{\rm shift}=0$ and $-30,000$~\kms.\label{fn1:EW}}
\footnotetext[3]{Measured from the total quasar population. Errors are based on number statistics.\label{fn1:balq}}
\end{minipage}
\end{table*}

\begin{figure*}
\includegraphics[width=174mm]{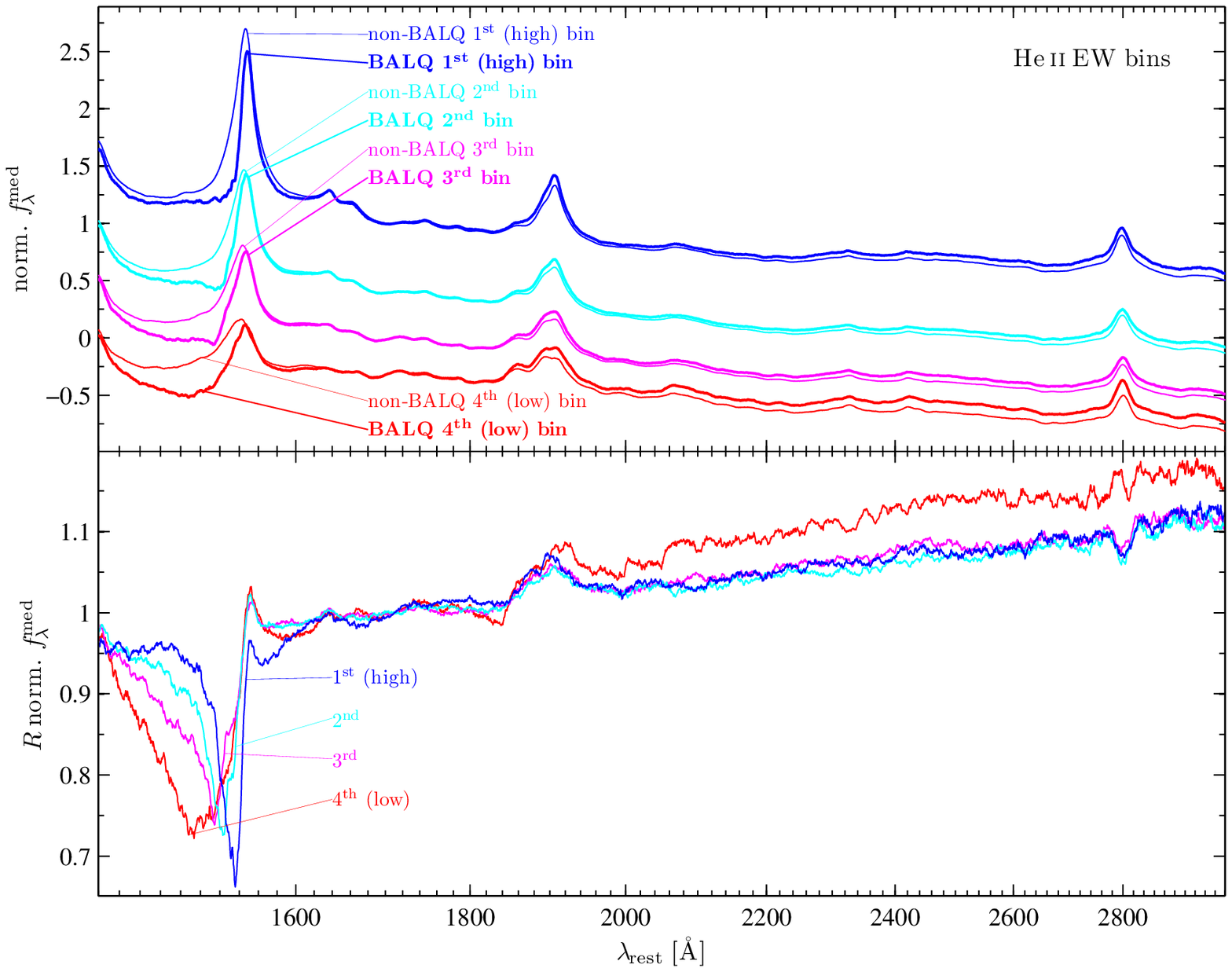}
\caption{Same as Fig.~\ref{fig:EW_subs_lt1700}, for the low-$z$ sample. The BALQ and non-BALQ samples are divided here to four subsamples. The 2nd to 4th bin spectra in the upper panel are shifted down for presentation purposes by 0.6, 1 and 1.3, respectively. The \CIV\ absorption profile becomes stronger and broader with lower \HeII\ EW values, as seen in the high-$z$ sample (Figure~\ref{fig:EW_subs_lt1700}).} \label{fig:EW_subs_ab1400}
\end{figure*}

Figure~\ref{fig:EW_subs_ab1400} presents the \HeII\ EW binning for the low-$z$ sample, this time binned to 4 bins (as further described below), allowed by the larger sample size (see Section~\ref{sec:methods}). The same trend of increasing depth and blueshift in the \CIV\ absorption profile with decreasing \HeII\ EW is observed here as well. There is also an excess emission in the $\CIII]+\SiIII]+\AlIII$ complex in the $R$ spectra, which does not depend on the \HeII\ EW. The excess is related to a trend with \auvl\ discussed below (Section~\ref{sec:trends_auv}). Table~\ref{tab:HeII_lowz} lists the bin median properties, including the median \MgII\ FWHM, $L(3000\mbox{\AA})$, \Mbh\ and \LLedd. It also presents the number of LoBALQs with similar \HeII\ EW. LoBALQs are found predominately in the lowest \HeII\ EW bin (43 out of 56 objects). In Appendix~\ref{sec:LoBALQ_comp} we present a LoBALQ composite based on the 56 objects we excluded from our low-$z$ sample in order to form the low-$z$ HiBALQ sample (Section~\ref{sec:methods}). The redder $R$ spectrum of the lowest \HeII\ EW bin, compared to the other three, might represent a smooth transition between HiBALQs and LoBALQs, where the latter have the lowest \HeII\ EW and the reddest $R$ spectrum (Appendix~\ref{sec:LoBALQ_comp}). The smooth transition scenario can be further explored by analysing the \MgII\ BAL dependence on the \HeII\ EW and reddening in LoBALQs. This requires a larger sample of LoBALQs and is outside the scope of this paper. 

Tables~\ref{tab:HeII_highz} and \ref{tab:HeII_lowz} list for each BALQ bin the observed BALQ fraction in DR7, which is defined as $\fbalq\equiv N_{\rm BALQs}/(N_{\rm BALQs}+N_{\rm non-BALQs})$. The tabulated errors on $\fbalq$ are minimal, and are based purely on number statistics. There is a strong trend between \fbalq\ and the \HeII\ EW. For the low-$z$ sample, \fbalq\ goes up from $7.0\pm0.4$ per cent for the highest \HeII\ EW bin, to $23\pm1$ per cent for the lowest bin. For the high-$z$ sample, a similar trend is found, and \fbalq\ goes up from $11\pm1$ per cent for the high \HeII\ EW bin to $22\pm2$ per cent for the low \HeII\ EW bin.

\begin{table*}
\begin{minipage}{180mm}
\caption{The median properties of \HeII\ EW binned low-$z$ objects.}\label{tab:HeII_lowz}
\begin{tabular}{@{}{l}*{12}{c}{c}@{}}
\hline
Class & $N_{\rm bin}$ & $N_{\rm obj}$ & \HeII &\auvl\fnrepeat{fn3:auv}& $\log$ &   \MgII& $\log$ & $\log$ &  $N_{\rm LoBALs}$\fnrepeat{fn3:LoBAL} & \multicolumn{3}{c}{\CIV\ BAL\fnrepeat{fn3:EW}} & \fbalq\fnrepeat{fn3:balq}\\
&  &  & EW & & $L(3000\mbox{\AA})$  &  FWHM& \Mbh & \LLedd &  & EW & $v_{\rm shift}^{\rm mean}$ & $\sigma$ & \\
& &  & [\AA] & & [erg s$^{-1}$]  & [\kms] & [M$_{\sun}$] & &  & [\AA] & [\kms] & [\kms] & [per cent]\\
\hline
BALQs & 1 & 399 & 7.0 &$-0.97$ & 45.90 & 3900 & 9.03 & $-0.48$ & 7 & 7.8 & $-4600$ & 2500& $7.0\pm0.4$\\
 & 2 & 399 & 4.7 & $-0.83$ & 45.97 & 4400 & 9.14 & $-0.52$  & 1 & 13.9 & $-9200$ & 5700 &  $10.3\pm0.5$\\
 & 3 & 399 & 3.0 & $-0.79$ & 46.00 & 4100 & 9.09 & $-0.44$ & 5& 15.7 & $-10900$ & 5900 & $14.1\pm0.8$\\
 & 4 & 399 & 0.9 & $-0.96$ &45.99 & 3400 & 8.93 & $-0.30$ & 43 &  22.5 & $-11900$& 6000 & $23\pm1$\\
\hline
non- & 1 & 5330 & 7.3 & $-0.77$ & 45.81 & 3800 & 8.93 & $-0.49$ \\
BALQs & 2 & 3461 & 4.7 & $-0.65$ & 45.93 & 4000 & 9.03 & $-0.47$ \\
& 3 & 2425 & 3.1 & $-0.61$ & 45.94 & 3700 & 8.98 & $-0.38$\\
& 4 & 1368 & 1.0 & $-0.69$ &45.93 & 3300 & 8.87 & $-0.30$\\
\hline
\end{tabular}
\footnotetext[1]{Measured between 1710 and 3000~\AA\ ($f_\nu\propto\nu^{\alpha}$).\label{fn3:auv}}
\footnotetext[2]{Number of LoBALQs (identified based on a detection of \MgII\ BAL; \citealt{shenetal11}) with \HeII\ EW within or nearest to the bin range.\label{fn3:LoBAL}}
\footnotetext[3]{Measured between $v_{\rm shift}=0$ and $-30,000$~\kms.\label{fn3:EW}}
\footnotetext[4]{Measured from the total quasar population. Errors are based on number statistics.\label{fn3:balq}}
\end{minipage}
\end{table*}

\subsection{\CIV\ BAL trends with $\alpha_{\rm UV}$}\label{sec:trends_auv}
\begin{figure*}
\includegraphics[width=174mm]{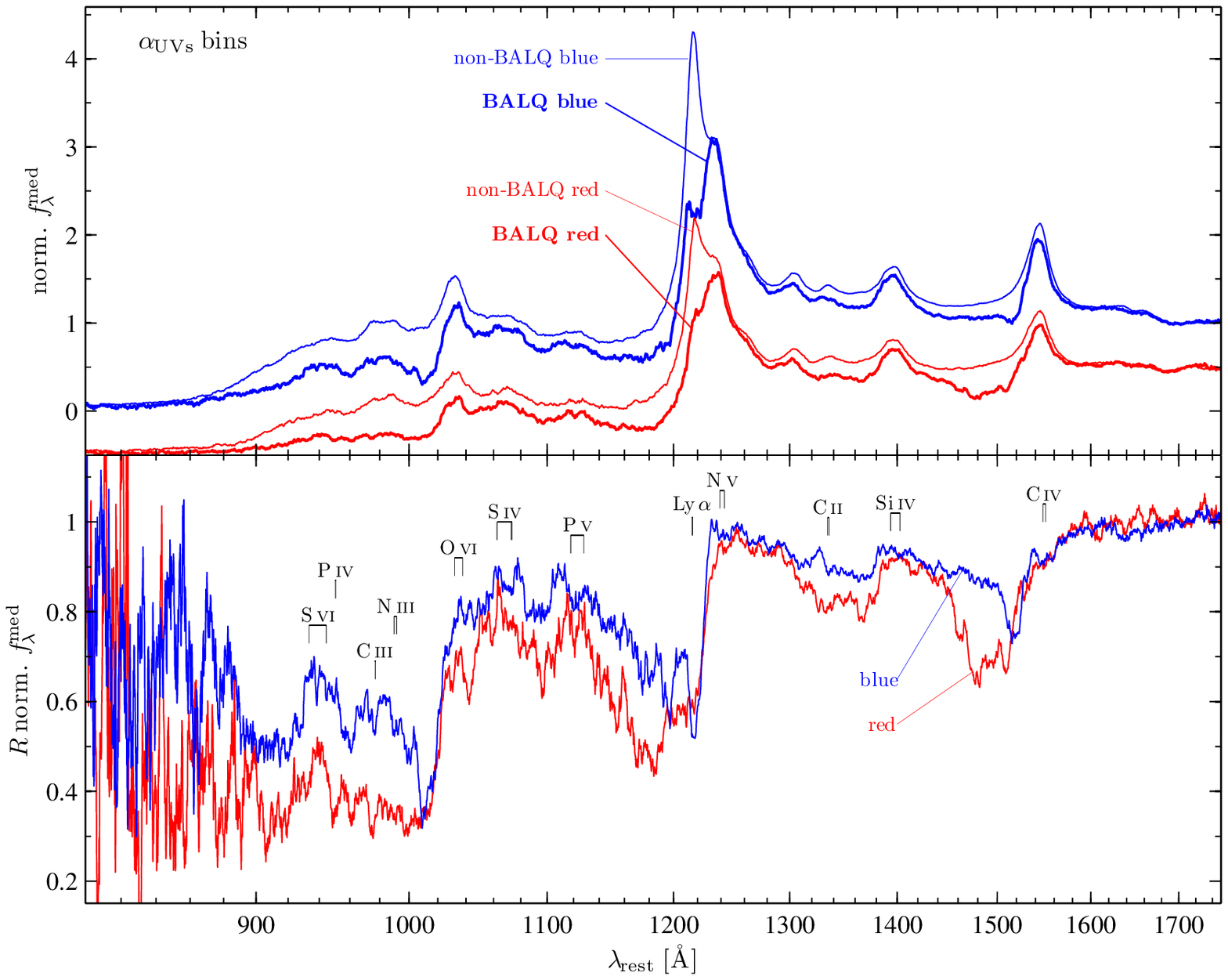}
\caption{The dependence of the intrinsic absorption on the \auvs\ for the high-$z$ sample. Top panel: the composite spectrum of BALQ and non-BALQ subsamples matched in \auvs\ range. The BALQ and non-BALQ red bins are shifted down by 0.5 for presentation purposes. Bottom panel: the ratio between the corresponding BALQ and non-BALQ composites. Note the stronger absorption for the redder \auvs. There is a residual reddening of the $R$ spectrum, since the median slope of the BALQ bin is redder than the slope of matched non-BALQ bin.}\label{fig:auv_subs_lt1700}
\end{figure*}

\begin{figure*}
\includegraphics[width=174mm]{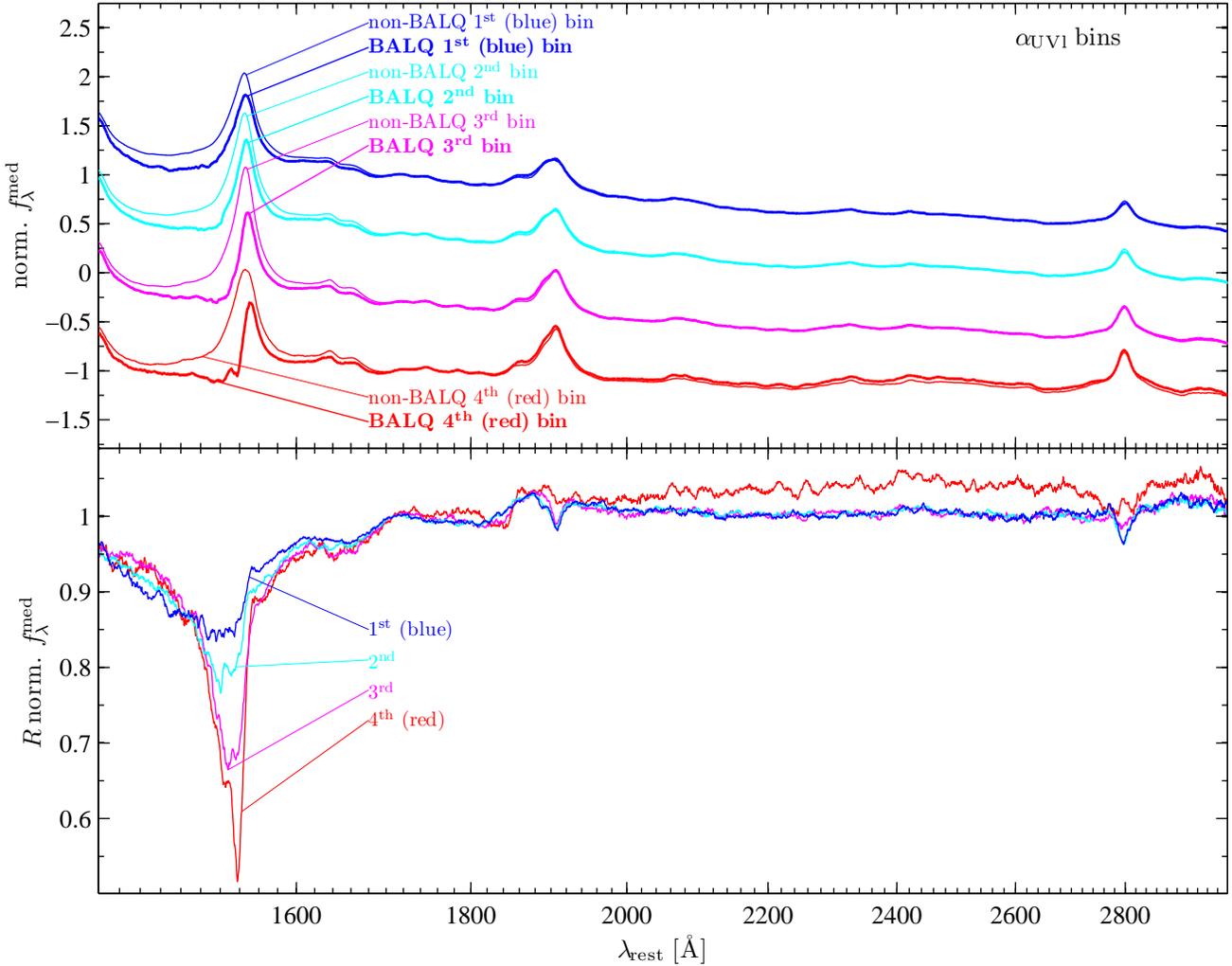}
\caption{The dependence of the intrinsic absorption on \auvl\ for the low-$z$ sample. The BALQ and non-BALQ samples are divided into four bins. The 2nd to 4th bin spectra are shifted down for presentation purposes by 0.6, 1.3 and 2, respectively. The \CIV\ absorption profile becomes deeper as \auvl\ becomes redder at the lower velocities, and remains unaffected at the highest velocities.} \label{fig:auv_subs_ab1400}
\end{figure*}

Figures~\ref{fig:auv_subs_lt1700} and \ref{fig:auv_subs_ab1400} explore the dependence of the BALQ absorption on $\alpha_{\rm UV}$. Figure~\ref{fig:auv_subs_lt1700} presents the high-$z$ sample binned into two \auvs\ bins. The \CIV\ absorption profile follows the trend observed in the \HeII\ EW binning, where the redder \auvs\ bin corresponds to the lower \HeII\ EW bin. Table~\ref{tab:auv_highz} lists the bin median properties of the high-$z$ sample. The median \auvs\ of the BALQ bins is redder than the slope of the matched non-BALQ bins. Although the bins are matched by the \auvs\ range (Section~\ref{sec:methods}), the distribution of \auvs\ values within the bin is different for the BALQs and non-BALQs. This slope difference yields the observed residual reddening of the $R$ spectrum (Fig.~\ref{fig:auv_subs_lt1700}, bottom panel).  Figure~\ref{fig:auv_subs_ab1400} presents the low-$z$ sample divided into four \auvl\ bins. The trend with \auvl\ is different from the trend with \auvs. The low velocity \CIV\ absorption profile becomes deeper as \auvl\ becomes redder, while at higher velocities the \CIV\ absorption profile remains unchanged. Table~\ref{tab:auv_lowz} summarizes the bin median properties. Note that the $\CIII]+\SiIII]+\AlIII$ complex emission becomes stronger with redder \auvl\ (Fig.~\ref{fig:auv_subs_ab1400}, top panel), and the BALQ and non-BALQ matched composites have a similar complex emission.

The fraction of BALQs increases as the objects get redder. At the high-$z$ sample, \fbalq\ increases from $11\pm1$ to $26\pm3$ per cent, while at the low-$z$ sample, \fbalq\ increases from $6.9\pm0.4$ to $21\pm1$ per cent from the bluest to the reddest quartile. Most of the detected LoBALQs (44 out of 56) have slopes at least as red as the reddest bin, as expected (see Section~\ref{sec:intro}).

\begin{table*}
\begin{minipage}{110mm}
\caption{The median properties of \auvs\ binned high-$z$ objects.\fnrepeat{fn2:auv}}\label{tab:auv_highz}
\begin{tabular}{@{}{l}*{7}{c}{c}@{}}
\hline
Class & Bin & $N_{\rm obj}$ & \auvs &\HeII & \multicolumn{3}{c}{\CIV\ BAL\fnrepeat{fn2:EW}} & \fbalq\fnrepeat{fn2:balq}\\
&  &  & & EW& EW & $v_{\rm shift}^{\rm mean}$ & $\sigma$ \\
& &  & &[\AA] & [\AA] & [\kms] & [\kms] & [per cent]\\
\hline
BALQs & blue & 100 & $-0.92$ &4.8 & 7.9 & $-$11700 & 6300 & $11\pm1$\\
 & red & 100 & $-1.90$ & 1.2 & 17.2 & $-$12600 & 4800 & $26\pm3$\\
\hline
non- & blue & 848 & $-0.72$ & 5.3\\
BALQs & red & 282 & $-1.66$ & 1.2\\
\hline
\end{tabular}
\footnotetext[1]{Measured between 1280 and 1710~\AA\ ($f_\nu\propto\nu^{\alpha}$).\label{fn2:auv}}
\footnotetext[2]{Measured between $v_{\rm shift}=0$ and $-30,000$~\kms.\label{fn2:EW}}
\footnotetext[3]{Measured from the total quasar population. Errors are based on number statistics.\label{fn2:balq}}
\end{minipage}
\end{table*}

\begin{table*}
\begin{minipage}{180mm}
\caption{The median properties of \auvl\ binned low-$z$ objects.\fnrepeat{fn4:auv}}\label{tab:auv_lowz}
\begin{tabular}{@{}{l}*{12}{c}{c}@{}}
\hline
Class & $N_{\rm bin}$ & $N_{\rm obj}$ & \auvl&\HeII & $\log$ & \MgII & $\log$ & $\log$ &  $N_{\rm LoBALs}$\fnrepeat{fn4:LoBAL} & \multicolumn{3}{c}{\CIV\ BAL\fnrepeat{fn4:EW}} & \fbalq\fnrepeat{fn4:balq}\\
&  &  & &EW & $L(3000\mbox{\AA})$  &  FWHM& \Mbh & \LLedd &  & EW & $v_{\rm shift}^{\rm mean}$ & $\sigma$ & \\
& &  & &[\AA]& [erg s$^{-1}$] & [\kms]  & [M$_{\sun}$] & &  & [\AA] & [\kms] & [\kms] & [per cent]\\
\hline
BALQs & 1 & 399 & $-0.48$ &3.5 & 45.92 & 4100 & 9.08 & $-0.49$ & 3 & 7.6& $-10900$ & 6200 & $6.9\pm0.4$\\
 & 2 & 399 & $-0.76$ & 4.0& 45.97 & 4100 & 9.09 & $-0.45$ & 3 & 8.2 & $-9000$ & 5600 & $10.7\pm0.6$\\
 & 3 & 399 & $-1.05$ &4.3 & 45.97 & 3900 & 9.05 & $-0.41$ & 6 & 6.8 & $-4700$ & 2300 & $14.5\pm0.8$\\
 & 4 & 399 & $-1.51$ &3.5 & 46.00 & 3600 & 9.02 & $-0.38$& 44 & 10.3 & $-4500$ & 2500 & $21\pm1$\\
\hline
non- & 1 & 5414 & $-0.44$ &4.6 & 45.86 & 3800 & 8.96 & $-0.46$ \\
BALQs & 2 &  3334 & $-0.76$ &5.4 & 45.92 & 3800 & 8.99 & $-0.43$\\
& 3 & 2349 & $-1.02$ &5.9 & 45.89 & 3700 & 8.97 & $-0.42$ \\
& 4 & 1526 & $-1.45$ &5.1 & 45.86 & 3600 & 8.92 & $-0.41$\\
\hline
\end{tabular}
\footnotetext[1]{Measured between 1710 and 3000~\AA\ ($f_\nu\propto\nu^{\alpha}$).\label{fn4:auv}}
\footnotetext[2]{Number of LoBALQs (identified based on a detection of \MgII\ BAL; \citealt{shenetal11}) with \auvl\ within or nearest to the bin range.\label{fn4:LoBAL}}
\footnotetext[3]{Measured between $v_{\rm shift}=0$ and $-30,000$~\kms.\label{fn4:EW}}
\footnotetext[4]{Measured from the total quasar population. Errors are based on number statistics.\label{fn4:balq}}
\end{minipage}
\end{table*}

\subsection{\CIV\ BAL trends with both \HeII\ EW and $\alpha_{\rm UV}$}\label{sec:trends_both}
\begin{figure}
\includegraphics[width=84mm]{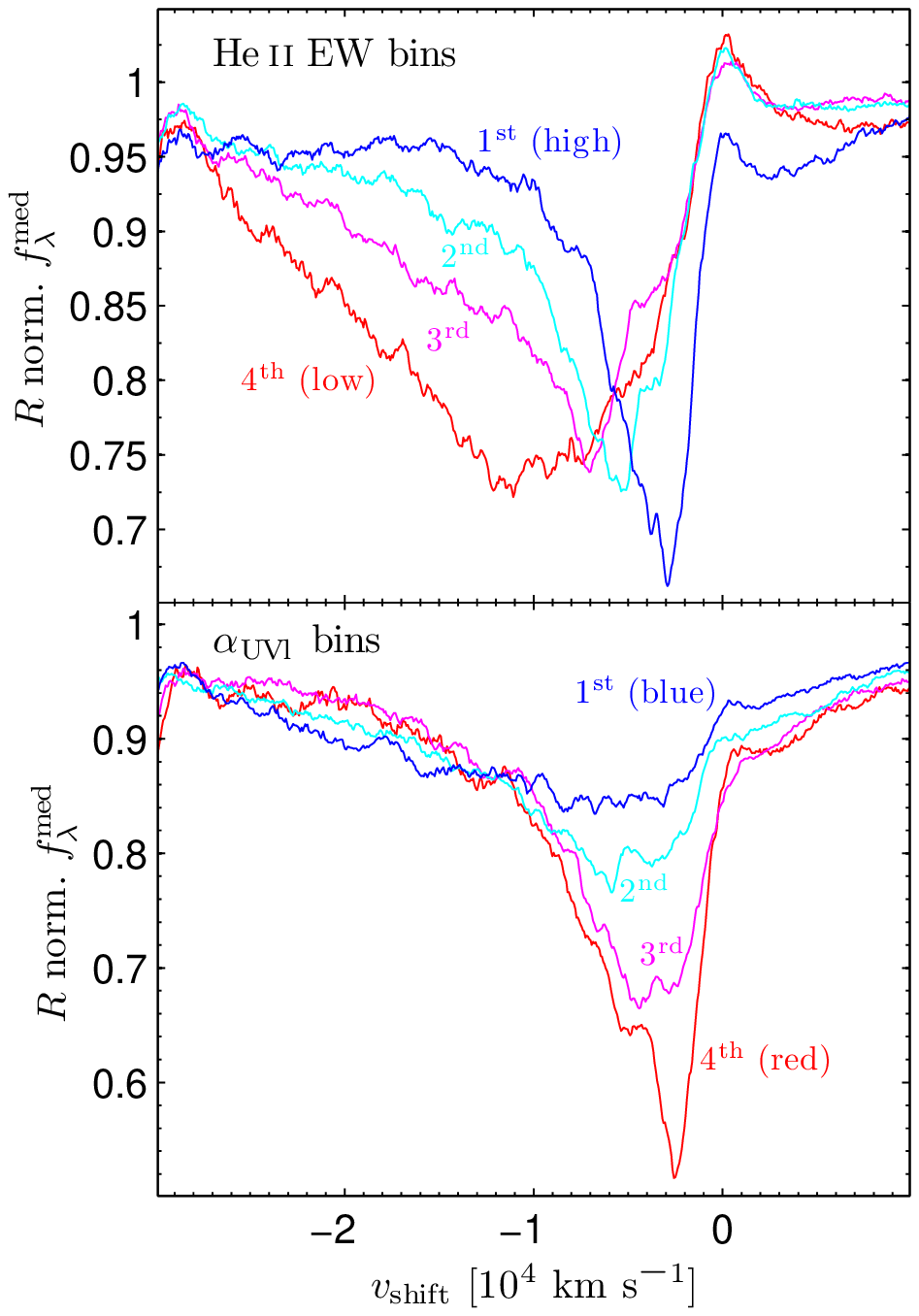}
\caption{A zoom in on the \CIV\ BAL profile for the composite ratios presented in Figs.~\ref{fig:EW_subs_ab1400} and \ref{fig:auv_subs_ab1400}. The absorption profile becomes broader and shifts to higher velocities as the \HeII\ EW becomes lower, and the absorption becomes deeper at $\vshift>-10,000$~\kms\ as \auvl\ becomes redder.}\label{fig:zoom_CIV_abs}
\end{figure}

Figure~\ref{fig:zoom_CIV_abs} shows a zoom-in on the \CIV\ absorption profile for the low-$z$ sample, and highlights the strong and different trends between the \CIV\ BAL properties and the \HeII\ EW and \auvl\ (not true for \auvs). The \HeII\ EW mostly controls the characteristic absorption \vshift, while \auvl\ mostly controls the absorption depth, in particular at $\vshift<10,000$~\kms. Thus, it appears that the \HeII\ EW and \auvl\ are independent parameters which span the \CIV\ absorption profile properties. Their independence can also be inferred from Table~\ref{tab:HeII_lowz}, which shows similar median \auvl\ values in the different \HeII\ EW bins, and similarly in Table~\ref{tab:auv_lowz}, where the different \auvl\ bins show similar median \HeII\ EW values. This is in contrast with the high-$z$ sample, where \auvs\ and the \HeII\ EW are correlated (Tables~\ref{tab:HeII_highz} and \ref{tab:auv_highz}).

The highest \HeII\ EW and the bluest \auvl\ bins may be incomplete in other BALQ studies based on a BI selection. A value of $\textrm{BI}>0$ requires a BAL with a significant absorption ($>10$ per cent) at least down to $\vshift=-5000$~\kms\ (given a minimal width of 2000~\kms, starting at $\vshift=-3000$~\kms, \citealt{weymannetal91}). Since we find that $\vshift$ decreases with increasing \HeII\ EW, the highest \HeII\ EW bin might be missing yet lower $\vshift$ BALQs, with peak absorption at $\vshift>-3000$~\kms. Similarly, BALQs residing in the bluest \auvl\ bin may be missed, since the peak-absorption is $\la10$ per cent for this bin (Fig.~\ref{fig:zoom_CIV_abs}). Our sample is derived based on a BI$_0$ and visual selection (see Section~\ref{sec:methods}), where BI$_0>0$ requires significant absorption ($>10$ per cent) at least down to $\vshift=-2000$~\kms. Thus, large \HeII\ EW, low \vshift\ BALQs are probably not missing in our sample, but weak BALs ($\la5$ per cent) possibly present in BALQs with the bluest \auvl\ may be missing.

\begin{figure*}
\includegraphics[width=174mm]{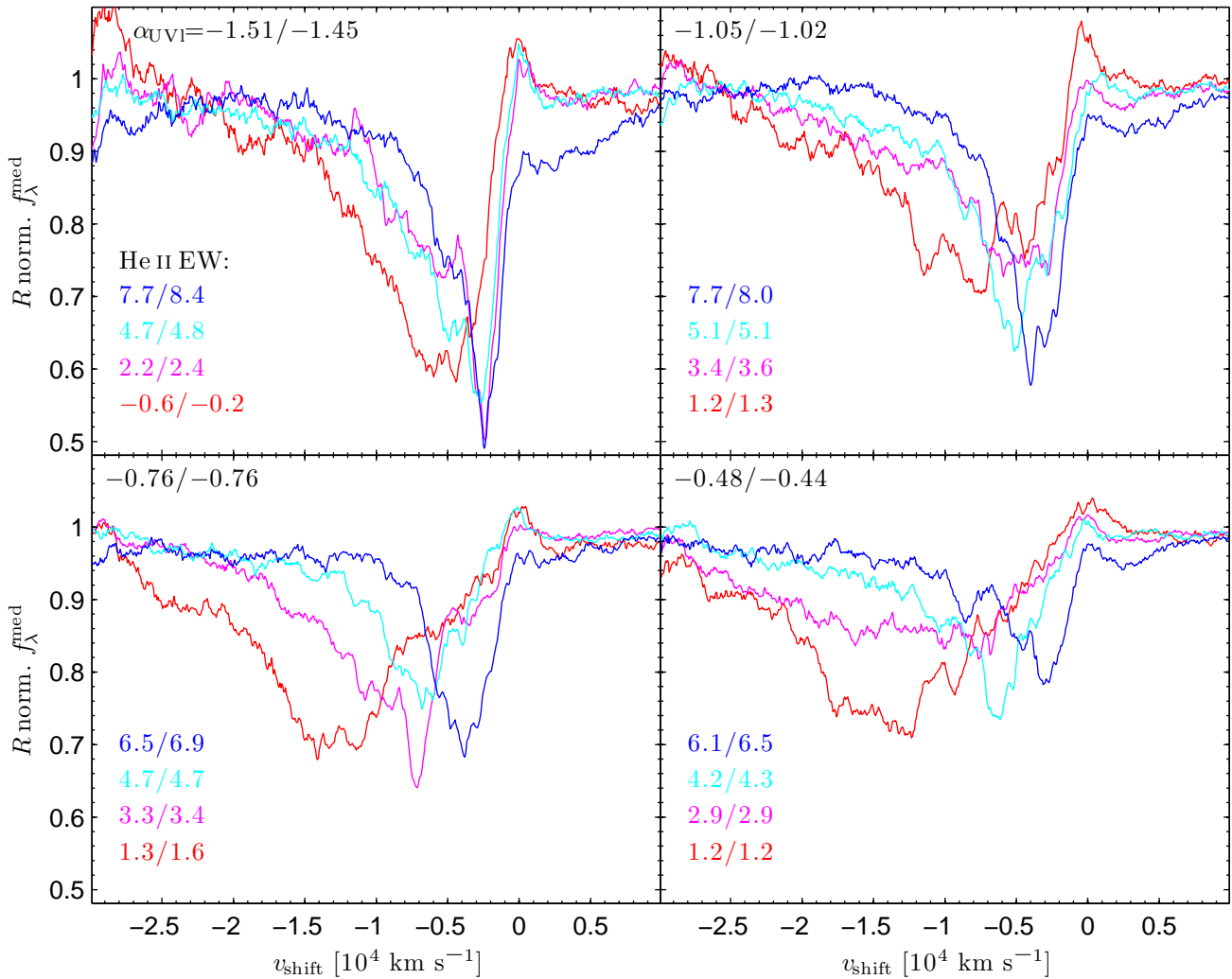}
\caption{The dependence of the \CIV\ BAL profile on \auvl\ and the \HeII\ EW. The BALQ and non-BALQ samples are first binned into \auvl\ bins (same as in Fig.~\ref{fig:auv_subs_ab1400}), and then each \auvl\ bin is binned based on the \HeII\ EW. The upper left corner in each panel provides the median \auvl\ values for the BALQ/non-BALQ bins used to generate the plotted ratio. The corresponding values for the four \HeII\ EW bins (in \AA) are indicated in the lower left corner of each panel. Note that all panels have the same y-scale. The \CIV\ BAL profiles become shallower as \auvl\ becomes bluer, and at each \auvl\ bin \vshift\ increases as the \HeII\ EW decreases. The span in \vshift\ increases as \auvl\ becomes bluer.} \label{fig:auv_HeII_subs_ab1400}
\end{figure*}

Figure~\ref{fig:auv_HeII_subs_ab1400} explores the \CIV\ BAL dependence on the \HeII\ EW for different values of \auvl. We divide the low-$z$ sample into $4\times4$ bins based on \auvl\ and the \HeII\ EW. First, the BALQ sample is divided into four \auvl\ `parent' bins, and matched non-BALQ bins are constructed (same procedure as in Fig.~\ref{fig:auv_subs_ab1400}). Then, each \auvl\ parent bin is divided into four \HeII\ EW bins, with matching non-BALQs from the corresponding \auvl\ non-BALQ bin. As Fig.~\ref{fig:auv_HeII_subs_ab1400} shows, the \CIV\ absorption at a given \auvl, shifts to higher \vshift\ as the \HeII\ EW decreases, at all four \auvl\ bins. Also, the maximum absorption depth at all \HeII\ EW bins, gets larger as \auvl\ gets redder. The change in \vshift\ is most prominent for the bluest two bins. Table~\ref{tab:auv_and_HeII_lowz} lists the median properties for the $4\times4$ bins. Note that the reddest and lowest \HeII\ EW BALQ and non-BALQ bins have median \HeII\ EW $<0$. This is an artefact which results from measuring the EW by using the continuum at 1700--1720~\AA, where in the reddest and weakest \HeII\ objects the flux density at \HeII\ is lower than at 1700--1720~\AA, leading to a negative EW (Section~\ref{sec:methods}). The negative \HeII\ EW has no effect on the binnings, which are independent of the absolute \HeII\ EW values. The value of \fbalq\ increases with decreasing \HeII\ EW, at a given \auvl, and also as \auvl\ gets redder, at a given \HeII\ EW. We find the smallest value of $\fbalq=4.2\pm0.4$ per cent in the highest \HeII\ EW and bluest \auvl\ bin, increasing to $31\pm4$ per cent in the lowest \HeII\ EW and reddest \auvl\ bin. Most of the LoBALQs (33/56) reside in the reddest and lowest \HeII\ EW bin.

\begin{table*}
\begin{minipage}{180mm}
\caption{The median properties of \auvl\ and the \HeII\ EW binned low-$z$ objects.\fnrepeat{fn5:auv}}\label{tab:auv_and_HeII_lowz}
\begin{tabular}{@{}{l}*{11}{c}{c}@{}}
\hline
Class/ & $N_{\rm bin}$ & $N_{\rm obj}$ & \HeII & $\log$ &  \MgII& $\log$ & $\log$ &  $N_{\rm LoBALs}$\fnrepeat{fn5:LoBAL} & \multicolumn{3}{c}{\CIV\ BAL\fnrepeat{fn5:EW}} & \fbalq\fnrepeat{fn5:balq}\\
\auvl&  &  & EW &  $L(3000\mbox{\AA})$ & FWHM & \Mbh & \LLedd &  & EW & $v_{\rm shift}^{\rm mean}$ & $\sigma$ & \\
& &  & [\AA] & [erg s$^{-1}$] &  [\kms]& [M$_{\sun}$] & &  & [\AA] & [\kms] & [\kms] & [per cent]\\
\hline
BALQs & 1 & 99 & 6.1 & 45.84 & 4500 & 9.08 & $-0.63$ & 2 & 6.2 & $-6000$ & 4200 & $4.2\pm0.4$\\
 $-0.48$ & 2 & 100 & 4.2 & 45.91 & 4700 & 9.17 & $-0.61$ & 0 & 12.2 & $-10200$ & 6000 & $6.3\pm0.6$\\
 & 3 & 100 & 2.9 & 45.98 & 3900 & 9.06 & $-0.42$ & 0 & 13.9 & $-13000$ & 6300 & $9.0\pm0.9$\\
 & 4 & 100 & 1.2 & 45.94 & 3300 & 8.90 & $-0.32$& 1 & 19.5 & $-13500$ & 5600 & $14\pm2$\\
\hline
non- & 1 & 2276 & 6.5 & 45.77 & 3900 & 8.95 & $-0.54$  \\
BALQs & 2 &  1488 & 4.3& 45.91& 4000& 9.03& $-0.47$\\
$-0.44$ & 3 &  1015 & 2.9 & 45.93 & 3600 & 8.98 & $-0.37$\\
& 4 & 598 & 1.2 & 45.90 & 3200 & 8.84 & $-0.30$ \\
\hline\hline

BALQs & 1 & 99& 6.5& 45.90& 4600& 9.14& $-0.59$ & 0 & 6.3 & $-4200$& 1800& $6.1\pm0.6$\\
 $-0.76$ & 2 & 100& 4.7& 45.97& 4300& 9.12& $-0.51$& 0 & 13.0 & $-10500$& 6200& $11\pm1$\\
 & 3 & 100 & 3.3& 46.04& 4300& 9.16& $-0.50$& 1 & 15.9& $-10400$& 5300& $15\pm2$\\
 & 4 & 100& 1.3& 45.97& 3400& 8.93& $-0.29$& 2 & 23.7 & $-13700$& 6200& $24\pm3$\\
\hline
non- & 1 &  1518& 6.9& 45.88& 3800& 8.98& $-0.46$ \\
BALQs & 2 & 827& 4.7& 45.96& 4000& 9.05& $-0.44$ \\
$-0.76$ & 3 &  547& 3.4& 45.97& 3700& 8.99& $-0.38$\\
& 4 &  318 & 1.6& 45.97& 3400& 8.92& $-0.27$\\
\hline\hline

BALQs & 1 & 99& 7.7& 45.91& 3700& 8.95& $-0.43$& 0& 8.5& $-4200$& 2000& $8.2\pm0.9$\\
 $-1.05$ & 2 & 100& 5.1& 45.97& 4300& 9.18& $-0.51$& 0& 11.8& $-6200$& 3400& $16\pm2$\\
 & 3 & 100 & 3.4 & 46.02& 3900& 9.08& $-0.41$& 1& 14.2& $-8800$& 5400& $19\pm2$\\
 & 4 & 99\fnrepeat{fn5:obj1} & 1.2& 45.97& 3400& 8.95& $-0.31$& 5 & 22.1& $-10900$& 5900& $29\pm3$\\
\hline
non- & 1 &  1107& 8.0& 45.83& 3700& 8.93& $-0.44$ \\
BALQs & 2 & 539& 5.1& 45.95& 3900& 9.04& $-0.42$ \\
$-1.02$ & 3 & 414& 3.6& 45.95& 3800& 9.00& $-0.41$ \\
& 4 &  243& 1.3& 45.97& 3600& 8.95& $-0.31$\\
\hline\hline

BALQs & 1 & 99& 7.7& 45.93& 3300& 8.86& $-0.37$& 3 & 6.3& $-3100$& 1400& $14\pm1$\\
 $-1.51$ & 2 & 100& 4.7& 46.05& 4000& 9.09& $-0.46$& 3 & 18.3& $-6900$& 4700& $22\pm2$\\
 & 3 & 100& 2.2& 46.02& 3700& 9.05& $-0.37$& 5 & 14.6& $-5700$ & 3700& $25\pm3$\\
 & 4 & 100& $-0.6$& 46.03& 3700& 8.93& $-0.33$& 33 & 25.9& $-9600$& 5800& $31\pm4$\\
\hline
non- & 1 &  620& 8.4& 45.78& 3500& 8.85& $-0.44$ \\
BALQs & 2 &  362& 4.8& 45.91& 3900& 9.01& $-0.45$\\
$-1.45$ & 3 & 307& 2.4& 45.95& 3500& 8.93& $-0.37$ \\
& 4 & 224& $-0.2$& 45.91& 3400& 8.87& $-0.31$ \\
\hline
\end{tabular}
\footnotetext[1]{The slope is measured between 1710 and 3000~\AA\ ($f_\nu\propto\nu^{\alpha}$).\label{fn5:auv}}
\footnotetext[2]{Number of LoBALQs (identified based on a detection of \MgII\ BAL; \citealt{shenetal11}) with \auvl\ and the \HeII\ EW within or nearest to the bin range.\label{fn5:LoBAL}}
\footnotetext[3]{Measured between $v_{\rm shift}=0$ and $-30,000$~\kms.\label{fn5:EW}}
\footnotetext[4]{Measured using the total quasar population. Errors are based on number statistics.\label{fn5:balq}}
\footnotetext[5]{An additional BALQ has \HeII\ EW smaller by more than 1~\AA\ than the minimal \HeII\ EW of the corresponding non-BALQ bin, and is excluded from the BALQ bin.\label{fn5:obj1}}
\end{minipage}
\end{table*}

\subsection{\CIV\ BAL dependence on other emission properties}\label{sec:trends_other}
\begin{figure}
\includegraphics[width=84mm]{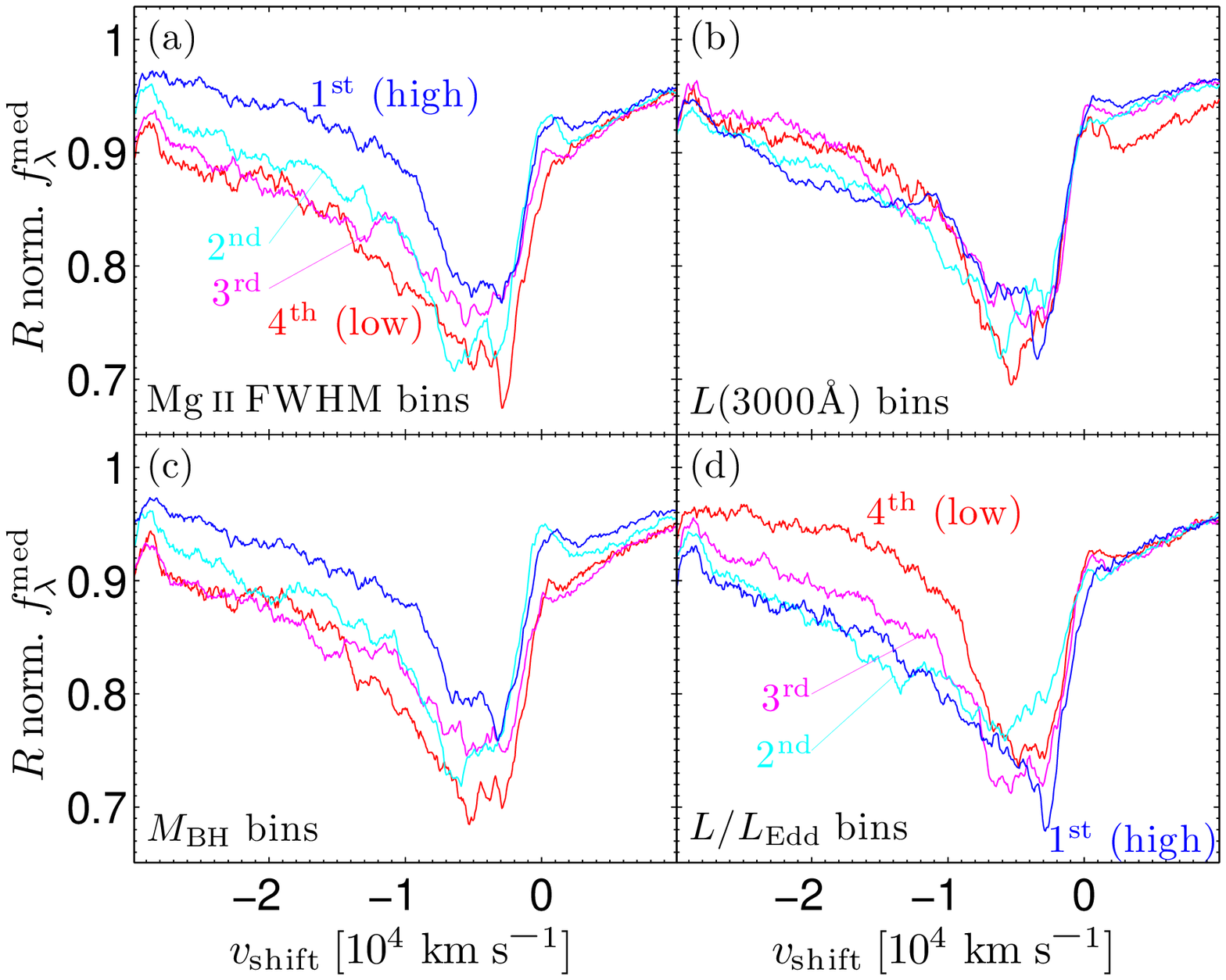}
\caption{Same as Fig.~\ref{fig:zoom_CIV_abs} but the samples are binned based on the \MgII\ FWHM, $L(3000\mbox{\AA})$, \Mbh\ and \LLedd. Since the $L(3000\mbox{\AA})$ range is small (see Table~\ref{tab:other_bins}), the trend in the \CIV\ absorption with \Mbh\ and \LLedd\ is a simple outcome of the correlation between the \CIV\ absorption and the \MgII\ FWHM [panel (a)]. Note that all presented trends are weaker than the trend between the \CIV\ BAL and the \HeII\ EW and \auvl\ (Fig.~\ref{fig:zoom_CIV_abs}).}\label{fig:zoom_CIV_phys_prop}
\end{figure}

Figure~\ref{fig:zoom_CIV_phys_prop} investigates whether a certain physical property underlies the above relations between the absorption profile and the \HeII\ EW and \auvl. The figure presents the \CIV\ BAL profile for samples binned based on the \MgII\ FWHM, $L(3000\mbox{\AA})$, \Mbh\ and \LLedd\ values. Table~\ref{tab:other_bins} lists the property median value of each bin for each of the 4 binning procedures. There is no trend in the absorption profile with $L(3000\mbox{\AA})$. There is a trend of stronger and broader absorption profile with decreasing \MgII\ FWHM, however the dynamical range in the absorption properties is significantly smaller than the range found for the \HeII\ EW or \auvl\ binning. Similar trends are seen with \Mbh\ and \LLedd. However, these two parameters are strongly correlated with the \MgII\ FWHM, given the  small range in $L(3000\mbox{\AA})$ (Table~\ref{tab:other_bins}), and since $L(3000\mbox{\AA})$ displays no trend, the \CIV\ BAL trend with \Mbh\ and \LLedd\ is equivalent to the trend with \MgII\ FWHM.  Also, $f_{\rm BALQ}$ is either nearly constant, or shows a range of generally less than a factor of two, which is another indication these four parameters are not the fundamental parameters that underlie the \CIV\ BAL profile. Thus, the \HeII\ EW and \auvl\ appear to be the primary parameters. The possible underlying physical explanation is discussed below (Section~\ref{sec:phys}).

A cautionary note is that one should keep in mind that the evaluated trends with \Mbh\ and \LLedd\ may not appear significant, since the sample does not cover a large enough dynamical range in $L$ ($\ga1.5$~dex) to surpass the $\sim\pm0.5$~dex uncertainty in the \Mbh\ prescription calibration \citep{laor98, krolik01, bentzetal09, wooetal10}.

\begin{table*}
\begin{minipage}{180mm}
\caption{The property median value of \MgII\ FWHM, $L(3000\mbox{\AA})$, \Mbh\ and \LLedd\ binned low-$z$ samples.\fnrepeat{fn6:note}}\label{tab:other_bins}
\begin{tabular}{@{}{l}*{10}{c}{r}*{2}{c}@{}}
\hline
Class & $N_{\rm bin}$ & \multicolumn{3}{c}{\MgII\ FWHM} & \multicolumn{3}{c}{$\log L(3000\mbox{\AA})$} & \multicolumn{3}{c}{$\log \Mbh$} & \multicolumn{3}{c}{$\log \LLedd$}\\
& &  [\kms]  & $N_{\rm obj}$ &\fbalq  & [erg s$^{-1}$] & $N_{\rm obj}$ & \fbalq & [M$_{\sun}$]  & $N_{\rm obj}$ & \fbalq & &$N_{\rm obj}$ & \fbalq\\
\hline
BALQs & 1  &6200&399& $13.5\pm0.7$ & 46.26& 399 & $15.2\pm0.8$&9.47&399& $15.5\pm0.8$&0.07&399& $12.1\pm0.6$\\
 & 2  &4500&399& $11.2\pm0.6$ & 46.04 & 399& $13.9\pm0.7$ & 9.16&399& $12.2\pm0.6$ &$-0.31$&399& $10.6\pm0.6$\\
 & 3 &3400&399& $9.8\pm0.5$ &45.90 & 399& $11.9\pm0.6$ &8.91&399& $9.4\pm0.5$ &$-0.55$&399& $10.8\pm0.6$\\
 & 4 &2200&399& $11.1\pm0.6$ & 45.66 & 399& $7.4\pm0.4$ &8.49&399& $9.7\pm0.5$ &$-0.90$&399& $11.5\pm0.6$\\
\hline
non- & 1  &6200 & 2567 && 46.26 & 2225 &&9.47&2167&&0.03&2889\\
BALQs & 2 & 4400 &3168& &46.04& 2469 &&9.17&2876&&$-0.31$&3350\\
& 3 &3400&3693&& 45.89& 2950&&8.92&3861&&$-0.55$&3312\\
& 4 &2200&3187 &&45.60 & 4981&&8.51&3732&&$-0.91$&3084\\
\hline
\end{tabular}
\footnotetext[1]{The property median value is followed by the number of objects in the bin and the observed fraction of BALQs from the total quasar population (in per cent).\label{fn6:note}}
\end{minipage}
\end{table*}

\subsection{Comparison with previous studies}\label{sec:compr}
The results of this study are consistent with previous findings. \citet{richardsetal02} and \citet{reichardetal03} report that BALQs have slightly weaker \HeII\ emission than non-BALQs. \citet{richardsetal02} find that their LoBALQ composite has a weaker \HeII\ emission than the HiBALQ and non-BALQ composites. This is consistent with most of the LoBALQs residing in our lowest \HeII\ EW bin. \citet{reichardetal03} also find the HiBALQs to be redder than non-BALQs, and the LoBALQs to be the reddest. \citet{richardsetal03} report a trend between the quasar colour and \fbalq\ by similarly dividing their sample into four bins. Their \fbalq\ goes up from $\sim3$ per cent for the bluest bin (compared to our $6.9\pm0.4$ per cent) to $\sim18$ per cent for the reddest bin ($21\pm1$ per cent). A relationship between reddening and the \HeII\ EW that breaks for the reddest bin (Table~\ref{tab:auv_lowz}) was also previously reported by \citet{richardsetal03}. The large fraction of LoBALQs found here in bins that have the largest \CIV\ absorption is consistent with \citet{allenetal11}, who find that LoBALQs reside mainly in objects with a strong and broad \CIV\ absorption (i.e., a high BI; see also Appendix~\ref{sec:LoBALQ_comp}). \citet{trumpetal06} report that BALQs have broader emission lines than non-BALQs, which is consistent with the high \MgII\ FWHM bin having the largest \fbalq\ (Table~\ref{tab:other_bins}). The small range in luminosity of the samples analysed here, prevents detection of the Baldwin effect (e.g., \citealt{gibsonetal09}). In this small range, we find a monotonic growth of \fbalq\ with $L(3000\mbox{\AA})$ (from $7.4\pm0.4$ to $15.2\pm0.8$ per cent), which is also reported by \citet{allenetal11} and attributed to a S/N selection effect (see also \citealt{kniggeetal08}). The trend found here is also affected by this effect, since there is a tight correlation in the low-$z$ sample between the spectral S/N and $L(3000\mbox{\AA})$, with the Spearman rank-order correlation coefficient of $r_{\rm S}=0.57$ and 0.66 for the BALQ and non-BALQ samples, respectively (null probability of $\log \textrm{Pr}<-130$). The median S/N per resolution element after smoothing (Section~\ref{sec:methods}) monotonically increases from $\la40$ for the lowest-$L$ bin objects to $\ga100$ for the highest-$L$ bin objects, and some of the absorption is shallow enough to be affected by the S/N level (see Section~\ref{sec:trends_both}).  

\section{A physical interpretation}\label{sec:phys}
We describe below a possible physical context for our findings. First, we suggest how the \HeII\ EW, which may be an indicator of the relative strength of the ionizing continuum beyond 4~Ryd, controls the \CIV\ BAL \vshift\ and \fbalq\ (Section~\ref{sec:phys_EW}). Then, we put forward a possible explanation on how \auvl, which may be a reddening indicator, controls the \CIV\ BAL depth and \fbalq\ (Section~\ref{sec:phys_auvl}). Finally, we briefly discuss connections to earlier studies (Section~\ref{sec:phys_other}).

\subsection{\CIV\ BAL trends with the \HeII\ EW}\label{sec:phys_EW}
\emph{What produces the strong trend between the \CIV\ BAL \vshift\ and the \HeII\ EW?} The \HeII\ $\lambda$1640 emission line is a recombination line, and provides a measure of the number of \HeII\ ionizing photons, i.e. photons above 54~eV, which are absorbed by the BLR gas. The \HeII\ EW serves as a measure of the relative strength of the EUV continuum above 54~eV, compared to the near UV continuum. As the BAL outflow increases its radial velocity, its density must drop, and its ionization parameter must rise, potentially leading to complete ionization. Complete ionization halts further acceleration of a radiation pressure driven outflow. A higher \HeII\ EW implies a higher ionization parameter at a given outflow density, and thus overionization of the outflow already at a higher density, i.e.\ at an earlier stage of the wind acceleration, where the outflow velocity is smaller. Even if the outflow is not radiation pressure driven (say it is magnetically driven), a harder ionizing continuum will overionize the outflow already at lower velocities, but due to different qualitative reasons \citep{fukumuraetal10a, fukumuraetal10b}. In the magnetically driven wind model of \citet{fukumuraetal10a}, the wind density and velocity go as $1/r$ and $1/\sqrt{r}$, respectively, where $r$ is the distance from the ionizing source. The density law implies that the ionization parameter also goes as $1/r$. Thus, in contrast with a radiation driven outflow, here the ionization parameter increases with increasing density and decreasing $r$ i.e., increasing velocity. However, for a harder ionizing continuum, the magnetically driven outflow gets overionized up to a larger $r$, i.e. down to a lower velocity. Thus, the \CIV\ BAL extends to lower maximal velocities, as expected for a radiation driven outflow, and as found in this study.

To prevent overionization, \citet{murrayetal95} suggested the presence of a shielding gas, which  blocks highly ionizing photons directed towards the outflow, preventing its overionization. Thus, ions such as \CIV\ remain, despite the rise in the ionization parameter with increasing velocity. The observed relationship between \vshift\ and the \HeII\ EW suggests that an outflow overionization does take place when the continuum which illuminates the BLR is hard enough. A large \vshift\ is obtained when the AGN ionizing continuum is soft, possibly enough to prevent overionization. Thus, one may not need to invoke a separate shield to prevent overionization. The suggestion that the observed weakness of the \HeII\ EW is sufficient to prevent overionization of the outflow, can be explored quantitatively through photoionization modelling of wind model.

The rise in \fbalq\ with decreasing \HeII\ EW is expected as a larger outflow volume is not overionized, leading to a larger CF of the BAL absorber. A magnetohydrodynamic wind model may also produce a similar rise in the CF, as a softer ionizing continuum implies absorbing wind which extends to a smaller $r$, and may produce a larger CF of the continuum source \citep{fukumuraetal10b}. It should be noted that since we are analysing average absorption profiles, the measured CF has an inherent degeneracy between the CF as seen by an observer, the global covering factor of BAL gas (i.e., the CF as seen by the continuum source) and the fraction of BALQs absorbing at a given \vshift.

Alternatively, the \HeII\ EW may indicate the fraction of the sky of the ionizing continuum covered by the BLR gas $\Omega$ instead of a measure of the strength of the EUV continuum above 54~eV. However, there are various relations favouring the interpretation of the \HeII\ EW as an ionizing SED indicator. The EUV slope is similar to $\alpha_{\rm ox}$ \citep{zhengetal97, laoretal97, telferetal02}. The $\alpha_{\rm ox}$ is correlated with luminosity \citep{stratevaetal05, steffenetal06, justetal07}, and luminosity only \citep{sl12}. Similarly, the \HeII\ $\lambda$4686 EW is correlated with luminosity only \citep{borgre92}. Thus, the ionizing continuum gets softer with increasing luminosity, leading to the drop in the \HeII\ EW with increasing luminosity. Also, lower luminosity objects have a flatter EUV slopes (\citealt{scottetal04}; cf.\ \citealt{telferetal02} and \citealt{shulletal12}). In addition, a BLR $\Omega$ trend with luminosity is not expected to directly affect \vshift. Future observations can explore directly whether the \HeII\ EW is correlated with the EUV slope.

In contrast with the \CIV\ BAL, the \NV\ and \OVI\ BALs do not show a prominent difference in their absorption profiles with the \HeII\ EW (Fig.~\ref{fig:EW_subs_lt1700}). The blue wing of \OVI\ may be affected by \NIII\ $\lambda$991 absorption, and to a lesser extent by \CIII\ $\lambda$997 absorption, while the \NV\ absorption is affected by \Lya\ absorption. Note also that the \NV\ and \OVI\ ions are produced and destroyed by photons at energies significantly above 54~eV, in contrast with the \CIV\ ion, where these energies (47.9--64.5~eV, Section~\ref{sec:mean_prop}) are just below and above 4~Ryd. One thus needs to further explore with photoionization calculations the relation between the \HeII\ emission EW, and the expected \NV\ and \OVI\ columns relative to the \CIV\ column, as a function of the ionizing spectral shape.

\subsection{\CIV\ BAL trends with \auvl}\label{sec:phys_auvl}
\emph{What produces the trend between the absorption depth and \auvl?} The value of \auvl\ may be interpreted as a viewing angle indicator, if it is affected by reddening, and if the dust tends to reside in the symmetry plane of the system. A support for this scenario is provided by a relation between the optical-UV slope (\aouv) and the degree of white light polarization, found by \citet{bl05} in the complete sample of \citet{borgre92} PG quasars. A redder \aouv\ is associated with a higher polarization, as expected for a system observed closer to edge on. \citet{bl05} also show that the change in \aouv\ is consistent with dust reddening (see also \citealt{sl12}, fig.~17 there). In this interpretation, if there is a clumpy and planar distribution of absorbers, then a redder \auv\ corresponds to viewing angles closer to edge-on, for which there is a larger probability that our line of sight intersects a UV absorbing gas cloud, and also a dusty gas cloud, i.e. a larger BAL CF and a larger reddening. This explains both the deeper absorption at a given \vshift, as a larger fraction of the central source is obscured, and the rise in \fbalq, as a larger fraction of lines of sight passes through an absorber. It is interesting to note that the median absorption depth increases from $\sim 5$ per cent at the shallowest \auvl, to $\sim 20$ per cent at the steepest \auvl\ (Fig.~\ref{fig:zoom_CIV_abs}), while \fbalq\ changes similarly from $6.9\pm0.4$ to $21\pm1$ per cent with \auvl\ (Table~\ref{tab:auv_lowz}).

BALQs observed at a larger inclination are also likely to have a larger absorbing column. However, since all lines are most likely saturated, they are not expected to show a rising absorbing EW. However, X-ray observations may reveal larger absorbing columns in steeper \auvl\ BALQs.

The inclination interpretation of \auvl\ implies that BALQs and non-BALQs, at a given \auvl, are observed at the same inclination. What separates BALQs from non-BALQs is then whether an outflow component happens to lie along our line of sight. This scenario requires an azimuthally asymmetric outflow structure, which is also hinted at by the absorption profile variability (e.g., \citealt{capellupoetal11, capellupoetal12}).

The inclination interpretation of \auvl\ can also explain the lower \vshift\ observed in redder BALQs (Figure~\ref{fig:auv_HeII_subs_ab1400}). Viewing a BALQ close to edge-on increases the probability that the line of sight goes through the base of outflowing absorbing gas, where the projected velocities are lower. A trend indeed observed in recent disc wind simulation by \citet[figure~5 there]{giupro12}. In contrast, the $\vshift>10,000$~\kms\ part of the outflow shows only a weak \auvl\ dependence (Figure~\ref{fig:zoom_CIV_abs}), which suggests the higher velocity outflow component is more spatially extended, and thus less inclination dependent.

We note in passing that binning by the shorter wavelengths \auvs\ leads to \CIV\ profiles which are very similar to those derived from the \HeII\ EW binning, which is in contrast with the result for the \auvl\ binning. Thus, \auvs\ is not orthogonal to the \HeII\ EW, as \auvl\ is. This can also be seen in Table~\ref{tab:auv_highz}, where the two \auvs\ bins show a factor of 4 difference in their median \HeII\ EW, in contrast with their similar values in the \auvl\ bins (Table~\ref{tab:auv_lowz}). Thus, \auvs\ appears to be related to the ionizing SED, as measured by the \HeII\ EW, while \auvl\ is independent of the ionizing SED, and likely provides a measure of the dust extinction (and indirectly the inclination). This qualitative difference is not unexpected, if the optical-UV continuum is produced by accretion disc emission. At long wavelengths, accretions discs are expected to show a universal SED slope, as we are observing the universal slope $L_\nu\propto\nu^{1/3}$ part of the disc emission (realistic model slopes are not that flat), which is independent of \Mbh\ and \LLedd. At short enough wavelengths, the slope starts to probe the position of the accretion disc spectral turnover region, and is thus a measure of the ionizing SED (e.g.\ \citealt{davlao11}).

\subsection{The luminosity dependence of $v_{\rm max}$}\label{sec:phys_other}
\citet{brandtetal00} and \citet{laobra02} found a clear correlation of the maximal outflow velocity $v_{\rm max}$ and the AGN luminosity in soft X-ray weak objects. This relation is quite robust as no Seyfert-level AGN is found to reach $v_{\rm max}\sim 10,000$~\kms, which is common in BALQs. A pure luminosity trend is simply explained by radiation pressure driven winds \citep[sec.~3.4 there]{laobra02}, which gives $v_{\rm max}\propto L^{1/4}$ for a wind launched from the BLR  with a constant force multiplier. The observed steeper relation $v_{\rm max}\propto L^{0.62\pm 0.08}$ can be interpreted as an indication of a force multiplier that rises with $L$ \citep{laobra02}. Such a rise may be caused by a softer ionizing SED at a higher $L$, which produces less overionization, and thus a rise in the force multiplier. A softer ionizing SED at a higher $L$ is consistent with the observed inverse relation between the \HeII\ EW and $L$. The range in AGN luminosity is too small here to test the luminosity dependence of $v_{\rm max}$ (but see tentative evidence in Appendix~\ref{sec:heii_L_bins}). However, the trend found here for \vshift\ with the \HeII\ EW, and the strong luminosity dependence of the \HeII\ EW, may indicate that the observed relation between $v_{\rm max}$ and the AGN luminosity found by \citet{laobra02}, may be partly driven by the \vshift\ versus \HeII\ EW relationship found here. A quick inspection of the data in \citet{laobra02} reveals that at a given luminosity, the highest $v_{\rm max}$ is reached by the lowest \HeII\ EW objects (which are also the soft X-ray weak quasars). However, a large sample which covers a large range in luminosity and the \HeII\ EW is required to clearly separate out the dependence of \vshift\ on $L$ and on the \HeII\ EW.

Inspection of Fig.~\ref{fig:EW_subs_ab1400} reveals that the \CIV\ emission profiles get weaker and blueshifted with decreasing \HeII\ EW, consistent with the finding of \citet{reichardetal03}. \citet[fig.~11 there]{richardsetal11} demonstrate the tight relation between the \HeII\ EW and the \CIV\ emission EW and asymmetry, which they interpret as an indication for emission from a wind component in the BLR (see also \citealt{leimoo04, kruczeketal11}). Above we found that the BAL wind component becomes more prominent as the EUV ionizing continuum gets weaker, which suppresses the overionization of the wind. The suggested BLR wind component may very well be related to the base of the BAL outflow, which is most likely fed by the BLR gas. As the wind accelerates from the BLR, its density and thus emissivity drops, but it remains visible as a BAL through its resonance line absorption. A weaker \HeII\ EW implies less overionization, which allows a higher wind emissivity closer to the base of the wind, producing a blushifted \CIV\ emission component. Unlike the blue wing of the \CIV\ emission that is probably produced by a matter-bounded wind component, the red wing of \CIV\ is probably produced by a non-wind (disc) component \citep{richardsetal11}, which is likely radiation bounded. As the relative strength of the EUV ionizing continuum becomes weaker, there is less production of C$^{3+}$ in the radiation-bounded disc component, which leads to a weaker emission in the \CIV\ red wing with decreasing \HeII\ EW, as observed.

\section{Can dust extinction explain the reddening?}\label{sec:dust}

The median BALQ SED is redder than the non-BALQ SED (Fig.~\ref{fig:all}). The difference between the two SEDs appears to be caused by the overall spectral slope, rather than local features. Is the difference in spectral slopes intrinsic to the illuminating source, or is it a result of extinction by foreground dust which is more common in BALQs?

\begin{figure*}
\includegraphics[width=174mm]{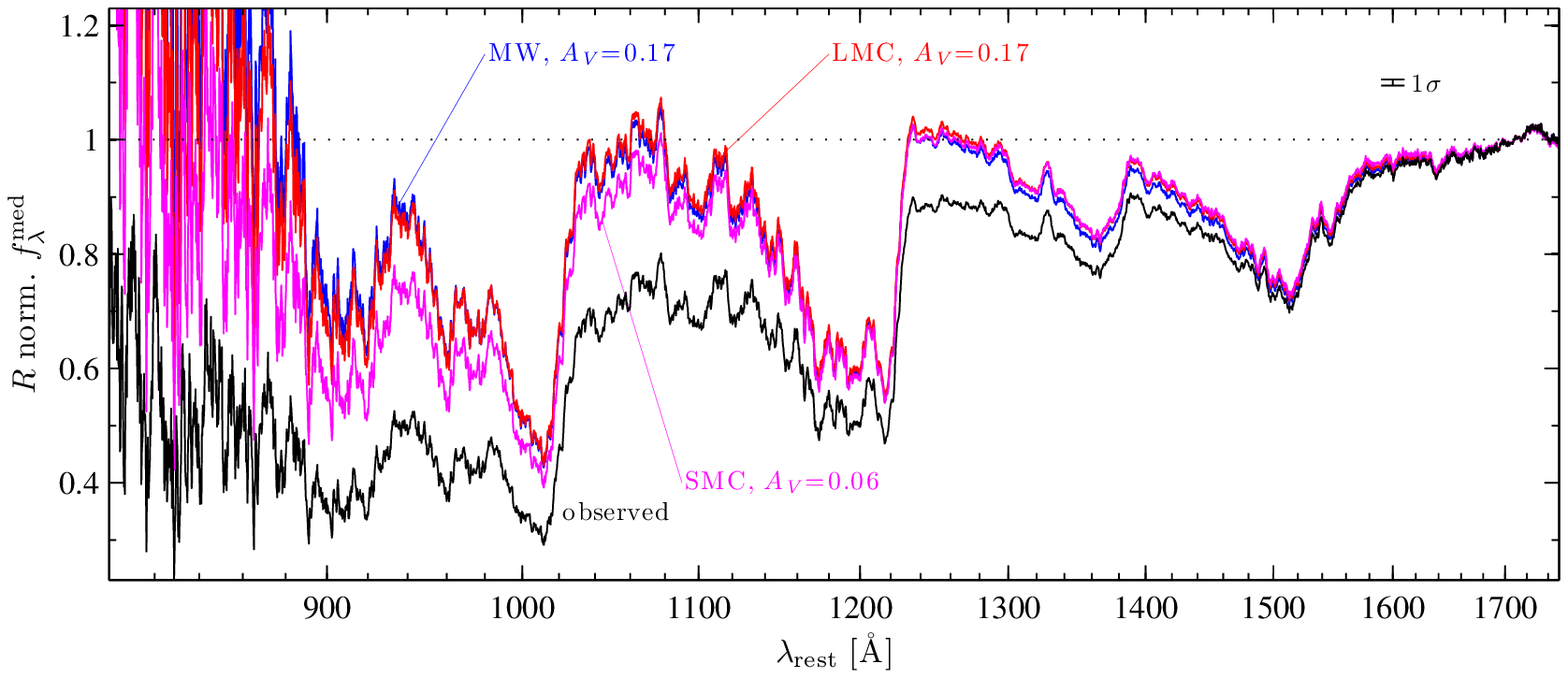}
\includegraphics[width=174mm]{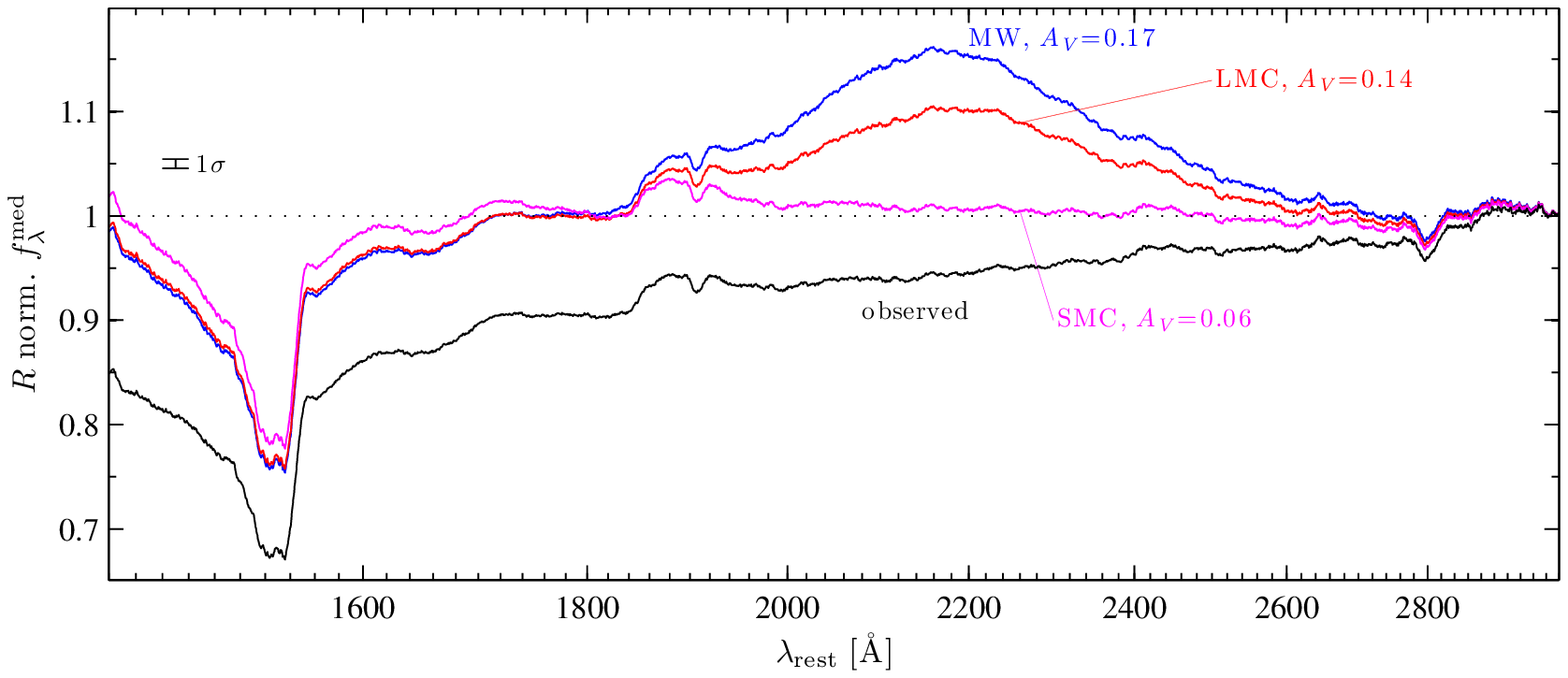}
\caption{High- and low-$z$ BALQ composites corrected for dust extinction assuming different extinction laws. The low-$z$ composite is normalized at 3000~\AA\ for presentation purposes. Three extinction-laws are adopted: MW ($R_{\it V}=3.2$) from \citet{cardellietal89}, and LMC ($R_{\it V}=3.4$) and SMC ($R_{\it V}=2.74$) from \citet{gordonetal03}. The best fit $A_{\it V}$ is estimated by eye inspection (see text). The error bars indicate the standard error in the median value. For the high-$z$ sample (upper panel), all extinction laws are possible, and the MW and LMC extinction corrections are essentially identical. However, the low-$z$ $R$ spectrum (bottom panel) excludes the MW and LMC extinction laws at a high significance level, as they produce a strong bump at $2000<\lambda_{\rm rest}<2400$~\AA, which is not observed. Note that both the low- and the high-$z$ composites are consistent with an SMC-like dust extinction with the same $A_{\it V}$ value of 0.06~mag.}\label{fig:dust}
\end{figure*}

Figure~\ref{fig:dust} presents corrections of the BALQ composite by several possible dust extinction laws. The BALQ composite is corrected to match the non-BALQ composite by adopting three types of dust extinction laws: Milky Way (MW; \citealt*{cardellietal89}), Large Magellanic Cloud (LMC) and Small Magellanic Cloud (SMC; the latter two laws are from \citealt{gordonetal03}).\footnote{All three extinction laws are reported only down to $\lamrest=1000$~\AA. Here we assume the analytic functions hold down to $\lamrest=800$~\AA. Given the low S/N of the $R$ spectrum at $\lamrest<1000$~\AA, the exact form of the extinction law used is not important.} The upper panel shows the de-reddening for the high-$z$ sample.  One cannot discriminate between the three extinction curves. The lower panel shows the de-reddening for the low-$z$ sample. Here one can clearly exclude a MW- and LMC-like extinction laws by $\sim13$ and 9 standard errors in the median $R$ at $\lamrest=2200$~\AA, as they produce a strong bump at $2000<\lamrest<2400$~\AA, which is not observed.\footnote{The standard error in the median $R$ is evaluated as follows. First, the standard deviation of $f_\lambda$ at 1700--1720~\AA\ is calculated for the median BALQ and non-BALQ spectra. Then, the error as a function of \lamrest\ is estimated by assuming the measured error is dominated by photon statistics i.e., the standard deviation is scaled by $\sqrt{f_\lambda(\lamrest)}$. Finally, the error in the $R$ spectrum is calculated by the standard error progression.}

A preference for SMC dust was found in various earlier studies of AGN. \citet{sprfol92} reach a similar conclusion for the reddening of LoBALQs compared to HiBALQs. \citet{glikmanetal12} find that the extinction of dust-reddened quasars in the \lamrest = 4000~\AA -- 2.4~$\mu$m range is best described by an SMC-like law (see also a brief overview of additional studies by \citealt*{pitmanetal00}). \citet{hopkinsetal04} report that the reddening of quasars in general is consistent with an SMC-like extinction law, while the MW and LMC laws are excluded. \citet{richardsetal03} also successfully fit dust-reddened non-BALQs assuming an SMC extinction law. Thus, the result derived here is clearly not new. However, these earlier studies are all based on the observed SED. The observed SED happens to show a shallow and broad dip at $\lambda\sim 2200$~\AA, produce by adjacent line emission, which can mimic some effect of the 2200~\AA\ extinction bump. The ratio plots used here eliminate all the intrinsic spectral features, and thus provide a strong constraint (to a level of $\sim 1$ per cent) on the possible extinction curve shape associated with the BAL outflow.

The SMC best-fit $A_{\it V}$ is evaluated \emph{independently} for the high- and low-$z$ samples. Both samples yield $A_{\it V}=0.06\pm0.01$~mag, where the range implies a deviation that is consistent with the standard error in the median $R$ spectrum. The best fit $A_{\it V}$ is estimated by eye inspection, requiring the non-BALQ and dereddened BALQ composites to match in wavelength regions unaffected by strong line emission and absorption, i.e.\ $\lamrest\approx1080$ and 1800~\AA\ for the high- and low-$z$ samples, respectively. The $A_{\it V}=0.06$~mag reddening fits the top three \HeII\ EW bins of the low-$z$ sample (Fig.~\ref{fig:EW_subs_ab1400}), while the lowest \HeII\ EW bin requires $A_{\it V}=0.09$~mag.

The $A_{\it V}$ values found here are mostly consistent with what was previously reported for BALQs \citep{reichardetal03, maddoxetal08, gibsonetal09}. \citet{allenetal11} find a trend between $A_{\it V}$ and $z$. The $A_{\it V}$ derived here is consistent with the value evaluated by \citet{allenetal11} for $1.6<z<2.0$ range (our low-$z$ sample), but it is $\sim3$ times smaller than the value for the $3.8<z<4.5$ range (our high-$z$ sample). We suspect the $3.8<z<4.5$ result of \citet{allenetal11} is an overestimate, as it over-corrects the BALQ \NV\ line emission (see fig.~22 there). 

The implied H column is $N({\rm H})\approx10^{20}$~\cmmt, using the $N({\rm H})=5.8\times10^{21}A_{\it V}/R_{\it V}$ relation \citep{draine11}, where $R_{\it V}=2.74$ for the SMC extinction law \citep{gordonetal03}. Given the above constraint on the \Ho\ column, based on the Lyman edge for a $\textrm{CF}=1$ absorber (Section~\ref{sec:const_ho}), the H column associated with the dust must be nearly fully ionized. If the gas is diffuse, the ionization can be maintained by the AGN on the host-galaxy scale. Since non-BALQs are also dust reddened (e.g., \citealt{richardsetal03,sl12}), the above estimate gives only the excess extinction in BALQs. To get the absolute extinction value, we evaluate the low-$z$ BALQ composite reddening relative to the bluest non-BALQ quartile composite, which is more likely to represent the intrinsic non-reddened SED. This yields $A_{\it V}=0.12$~mag and $N({\rm H})\approx3\times10^{20}$~\cmmt. A similar analysis for the high-$z$ sample, utilizing the bluest non-BALQ composite, yields $A_{\it V}=0.10$~mag and $N({\rm H})\approx2\times10^{20}$~\cmmt.

Since the SMC extinction curve is featureless, an intrinsically redder SED for BALQs cannot be excluded. The only support for the reddening interpretation is the remarkably good match, i.e., a ratio of unity for the dereddened spectra throughout the observed range, from 1000 to 3000~\AA.

\section{Conclusions}\label{sec:conclude}
We analyse the average absorption properties of high- and low-$z$ samples of BALQs from the SDSS DR7, which cover the wavelength range of 800--3000~\AA. The absorption properties are derived by taking the ratios of BALQ median spectra to the median spectra of matched samples of non-BALQs. We explore the absorption properties for a  number of subsamples selected based on various emission properties. We find the following.
\begin{enumerate}
\item No Lyman edge associated with the BAL absorbing gas is detected ($\tau<0.1$). Thus, on average, $\textrm{CF}\la0.1$ for a partially ionized absorber in BALQs.

\item The average absorption EW increases with the ionization state, from an absorption $\textrm{EW}\la1$\AA\ for \CII\ $\lambda$1335, to 5.8 for \SiIV, 9.3 for \CIV, 24.6 for \NV\ and 25.2~\AA\ for \OVI. This may indicate a rise in the covering factor of the BAL outflow with a rise in the ionization state.

\item The \HeII\ emission EW controls the typical \vshift\ of \CIV\ BAL, which increases from $\vshift< 7000$~\kms\ for $\textrm{EW}\ga6$~\AA\ to $\vshift> 15000$~\kms\ for $\textrm{EW}\la1$~\AA. The \HeII\ EW does not affect the absorption depth. The \HeII\ EW may indicate the strength of the EUV. A lower \HeII\ EW then implies a lower ionization of the outflow, which allows the outflow to reach higher velocities before being overionized. One may not need to invoke a shield to prevent over-ionization of the outflow, as over-ionization may be taking place in high \HeII\ EW objects, while in low \HeII\ EW objects a shield may not be needed.

\item The value of \auvl\ controls the \CIV\ peak absorption depth, which increases from $\sim0.15$ for $\auvl\approx-0.5$ to $\sim0.45$ for $\auvl\approx-1.5$, at $\vshift<10,000$~\kms. The value of \auvl\ does not affect the typical \vshift\ and the absorption profiles at $\vshift>10,000$~\kms. The value of \auvl\ may control the average inclination angle of the system, and thus the covering factor and absorbing column for a planar outflow, in particular at the lower velocities, which likely originate close to the base of the outflow.

\item The \HeII\ EW and \auvl\ also control the fraction of AGN which are BALQs. The fraction rises from $4.2\pm0.4$ per cent in blue ($\auvl\approx-0.5$) and strong \HeII\ ($\textrm{EW}\approx6$~\AA) AGN, to $31\pm4$ per cent in red ($\auvl\approx-1.5$) and weak \HeII\ ($\textrm{EW}\la 1$~\AA) AGN. Also, most of the LoBALQs ($\sim75$ per cent) have \HeII\ $\textrm{EW}\la1$~\AA. The \HeII\ EW may control the global CF of the BAL outflow, as seen by the ionizing source, as a lower \HeII\ EW allows the outflows to extend to larger scales before being overionized. On the other hand, \auvl\ may control the CF of absorbers along the line of sight, which increases with increasing inclination and the associated reddening.

\item The median SED of BALQs is consistent with excess reddening of purely SMC dust with $A_{\it V}=0.06$~mag compared to non-BALQs, or a reddening of $A_{\it V}=0.12$~mag compared to the bluest non-BALQs. Given the lack of associated \HI\ absorption, the dust is embedded in an ionized medium, possibly on the host-galaxy scale. LMC and MW dust are excluded by $\sim9$ and 13 standard deviation of the median value, respectively.
\end{enumerate}

The above interpretation of \auvl\ as a viewing angle indicator can be tested by looking for a trend between \auvl\ and the continuum polarization, in both BALQs and non-BALQs. One can also look for the expected relationship between \auvl\ and the X-ray absorbing column in BALQs. The relationships of \vshift\ and \fbalq\ with the \HeII\ EW, together with the inverse relationship of the \HeII\ EW with $L$, imply a steeper than linear rise in the kinetic wind luminosity with $L$, which may be relevant for feedback in the highest $L$ AGN.

\section*{Acknowledgments}
We thank G.\ T.\ Richards for many valuable comments. Fruitful discussions with N.\ Murray and E.\ Behar are acknowledged. We thank D.\ Kazanas and the anonymous referee for comments and suggestions. FH acknowledges support from the USA National Science Foundation grant AST-1009628. This research has made use of the Sloan Digital Sky Survey which is managed by the Astrophysical Research Consortium for the Participating Institutions; and of NASA's Astrophysics Data System Bibliographic Services.

\appendix

\section{The L\lowercase{o}BALQ \textbfit{R} spectrum}\label{sec:LoBALQ_comp}
As a sidenote, we use this opportunity to briefly present the LoBALQ median absorption properties based on
the low-$z$ BALQ sample. The LoBALQ $R$ spectrum is derived by forming a matched non-BALQ sample, which is matched by the \HeII\ EW distribution. Although most of the LoBALQs have small values of \HeII\ EW (the whole sample median EW is $-1.8$~\AA), several LoBALQs (13 out of 56) have somewhat larger \HeII\ EW ($>1$~\AA; Table~\ref{tab:HeII_lowz}), and the matching method used for HiBALQs (i.e., matching in the \HeII\ EW range) cannot be utilized here. To produce LoBALQ and non-BALQ composites with a matched \HeII\ EW distribution, we carry out the following procedure.
\begin{enumerate}
\item The LoBALQ \HeII\ EW distribution is calculated using 2~\AA\ bins (EW$_{\rm bin}$). At each EW$_{\rm bin}$, the number of LoBALQs with \HeII\ EW within the EW$_{\rm bin}$ range is divided by the total number of LoBALQs.
 
\item The non-BALQs are assigned to EW$_{\rm bin}$ based on their \HeII\ EW value. This yields the measured absolute non-BALQ distribution, $N_{\rm meas}(\textrm{EW}_{\rm bin})$.
 
\item The LoBALQ distribution is multiplied by the total number of non-BALQs. This yields the desired absolute non-BALQ distribution, $N_{\rm des}(\textrm{EW}_{\rm bin})$.
 
\item The non-BALQs are drawn for each EW$_{\rm bin}$ based on the following three criteria. If $N_{\rm meas}(\textrm{EW}_{\rm bin})=N_{\rm des}(\textrm{EW}_{\rm bin})$, then all non-BALQs in that EW$_{\rm bin}$ are drawn. If $N_{\rm meas}(\textrm{EW}_{\rm bin})< N_{\rm des}(\textrm{EW}_{\rm bin})$, then all non-BALQs are randomly drawn several times until the number of drawn objects equals $N_{\rm des}(\textrm{EW}_{\rm bin})$. If $N_{\rm meas}(\textrm{EW}_{\rm bin})>N_{\rm des}(\textrm{EW}_{\rm bin})$, then only $N_{\rm des}(\textrm{EW}_{\rm bin})$ of non-BALQs is randomly drawn.
 
 \item The median spectrum of the drawn non-BALQs is evaluated.
 
\item The resulting median non-BALQ spectrum depends on the particular non-BALQs that are randomly drawn. To eliminate this dependence, the procedure in items (iv) and (v) is repeated 300 times. A median of the 300 median spectra is adopted as the matched non-BALQ composite. 
\end{enumerate}
The above procedure produces a matched non-BALQ sample with a median \HeII\ EW of $-1.3$~\AA\ (compared to $-1.8$~\AA\ for LoBALQs). The LoBALQ and the matched non-BALQ composites have \auvl\ of $-1.99$ and $-1.38$, respectively. The resulting \CIV\ BAL EW, $v_{\rm shift}^{\rm mean}$ and $\sigma$ are 33.9~\AA, $-10,500$ and $5300$~\kms, respectively. 

Figure~\ref{fig:LoBALQ} presents the LoBALQ $R$ spectrum corrected for reddening (see below), where we also indicate the laboratory wavelength of \MgII\ $\lambda\lambda$2796.4, 2803.5, \AlIII\ $\lambda\lambda$1854.7, 1862.8 and several \FeII\ absorption lines following \citet[][table 3 there]{richards01}. Note the presence of prominent \FeII\ BALs in the LoBALQ $R$ spectrum (absorption $\textrm{EW}\ga7$~\AA). The LoBALQ \CIV\ BAL profile is consistent with the relationships, which are found for the low-$z$ HiBALQ sample, between \vshift\ and the \HeII\ EW, and between \CIV\ peak absorption and \auvl\ (Sections~\ref{sec:trends_EW} and \ref{sec:trends_auv}, respectively). The LoBALQ composite has a redder \auvl\ and a deeper \CIV\ peak absorption than the reddest HiBALQ bin ($\auvl=-1.99$ versus $-1.51$; compare to Fig.~\ref{fig:auv_subs_ab1400}). The \CIV\ BAL \vshift\ of LoBALQ composite is similar to \vshift\ of the lowest \HeII\ EW HiBALQ bin (Fig.~\ref{fig:LoBALQ}), although the LoBALQ median \HeII\ EW is lower (0.9 versus $-1.8$~\AA). The similarity in \vshift\ may be a result of the softer ionizing SED (i.e., lower \HeII\ EW for LoBALQs) being `too soft' to produce a \CIV\ absorber at higher \vshift. The higher \vshift\ absorber is probably found at larger distances from the ionizing source than a lower \vshift\ absorber, and the soft ionizing SED at those distances  results in an ionization parameter that is too low to produce \CIV\ absorption.

\begin{figure*}
\includegraphics[width=174mm]{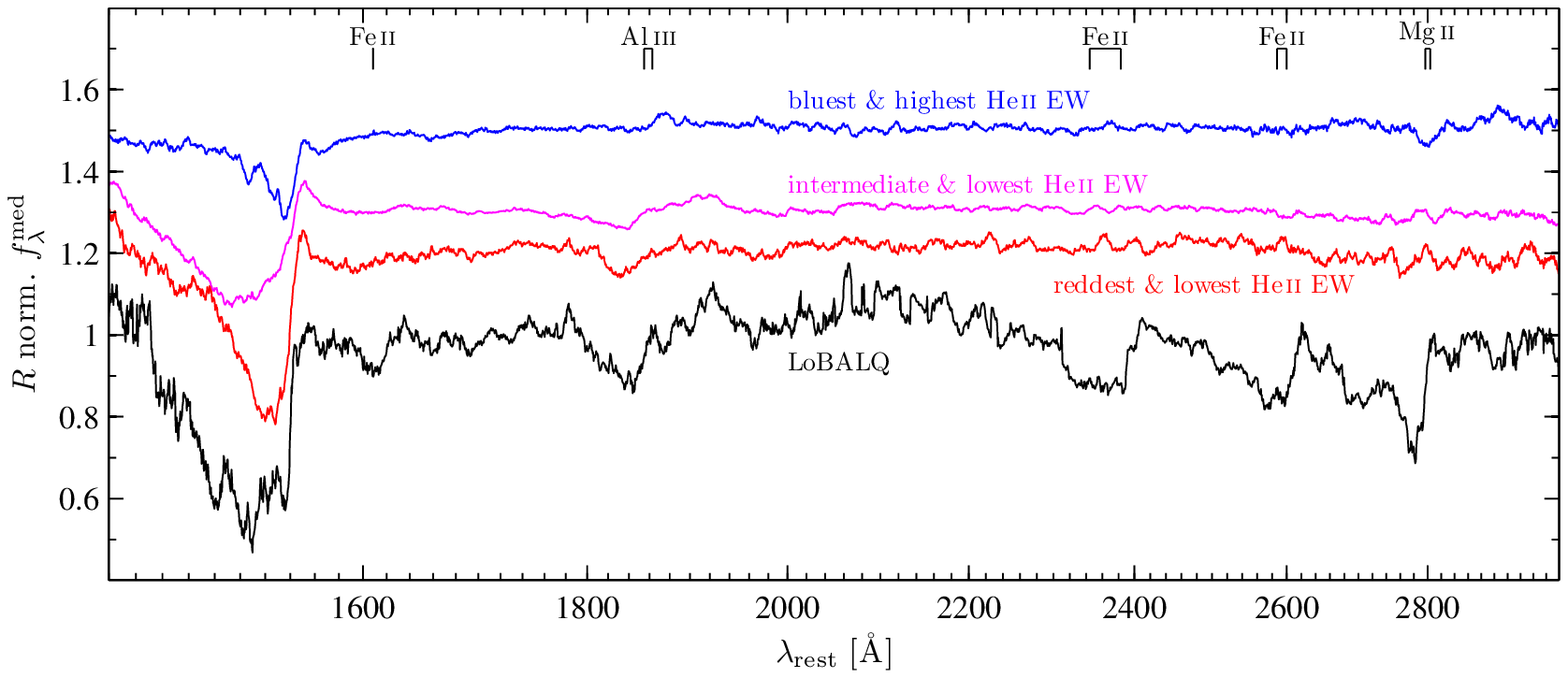}
\caption{The LoBALQ $R$ spectrum corrected for reddening (black line). The bluest and highest \HeII\ EW, the de-reddened lowest \HeII\ EW, and the reddest and lowest \HeII\ EW composites are presented for comparison (blue, magenta and red line, respectively), and are shifted up for presentation purposes by 0.5, 0.3, and 0.2, respectively. The LoBALQ $R$ spectrum is evaluated by using non-BALQs matched in the \HeII\ EW distribution (see text). Note that the de-reddened lowest \HeII\ EW bin has an intermediate slope relative to the other two HiBALQ bins. We indicate the laboratory wavelength of \MgII\ $\lambda\lambda$2796.4, 2803.5, \AlIII\ $\lambda\lambda$1854.7, 1862.8 and \FeII\ absorption lines, $\lambda\lambda$1608.5, 2344.2, 2382.8, 2586.7 and 2600.2. Note the prominent \FeII\ BAL troughs that are present in the LoBALQ $R$ spectrum. The LoBALQ sample extends the trend found for the low-$z$ HiBALQs between the \CIV\ absorption depth and \auvl\ (Fig.~\ref{fig:auv_subs_ab1400}). The LoBALQ sample has the reddest \auvl\ and the deepest \CIV\ absorption relative to HiBALQs. A comparison between  the different composites indicates that as the HiBALQ properties become more similar to those of LoBALQs (i.e., lower \HeII\ EW and redder \auvl), the HiBALQ absorption spectrum starts to resemble the LoBALQ spectrum. The \AlIII\ (\MgII) BAL strength increases from a non-detectable to prominent (marginal). The dip at the \MgII\ laboratory wavelength for the bluest and highest \HeII\ EW bin likely does not indicate absorption, since it results form a mismatch in \MgII\ peak line emission between BALQs and non-BALQs, and has no alignment with the LoBALQ \MgII\ BAL.}\label{fig:LoBALQ}
\end{figure*}

A comparison between LoBALQ and HiBALQ composites supports the scenario of a smooth transition between HiBALQs and LoBALQs, where the latter mostly occupy the reddest \auvl\ and the lowest \HeII\ EW parameter-space (Section~\ref{sec:trends_both}). Figure~\ref{fig:LoBALQ} presents the bluest and highest \HeII\ EW composite, the lowest \HeII\ EW composite (Fig.~\ref{fig:EW_subs_ab1400}) corrected for reddening (Section~\ref{sec:dust}), and the reddest and lowest \HeII\ EW composite. The first composite has EW = 6.1~\AA\ and $\auvl=-0.48$ (Table~\ref{tab:auv_and_HeII_lowz}), and its $R$ spectrum has no detectable \AlIII\ and \MgII\ BALs. The dip at \MgII\ laboratory wavelength results from a mismatch in \MgII\ peak line emission between the matched BALQs and non-BALQs, and likely does not indicate true absorption. The second composite has EW = 0.9~\AA\ and $\auvl=-0.96$ (Table~\ref{tab:HeII_lowz}), and has a marginal \AlIII\ BAL and non-detectable \MgII\ BAL. The third composite has a \HeII\ EW = $-0.6$~\AA\ and $\auvl=-1.51$ (Table~\ref{tab:auv_and_HeII_lowz}), and its $R$ spectrum has a prominent \AlIII\ BAL and a marginal \MgII\ BAL. There are additional evidence of \AlIII\ BAL strength dependence on both \HeII\ EW and \auvl. HiBALQ binning based on \HeII\ EW (\auvl) produces composites with similar \auvl\ (\HeII\ EW), but with different \AlIII\ BAL strength (Figs.~\ref{fig:EW_subs_ab1400} and \ref{fig:auv_subs_ab1400}). Note that the $R$ spectrum in the \MgII\ region in Figs.~\ref{fig:EW_subs_ab1400} and \ref{fig:auv_subs_ab1400} is dominated by a mismatch in the \MgII\ peak line-emission, which hinders a detection of any weak \MgII\ BAL. The \AlIII\ BAL appears first, as the ionization parameter becomes lower, since \MgII\ ions are destroyed by 15~eV photons, while \AlIII\ ions are produced by 19~eV photons. As the HiBALQ properties (i.e., \HeII\ EW and \auvl) start to resemble those of LoBALQs, low ionization BALs begin to appear in the HiBALQ spectrum, and it becomes more similar to the LoBALQ spectrum. This suggests LoBALQs are the extreme high inclination and soft ionizing SED members of the BALQ population, rather than a distinct population of AGN.

The reddening of the LoBALQ $R$ spectrum is consistent with an SMC dust extinction of $A_{\it V}=0.22$~mag i.e., $\textit{E(B$-$V)}=0.08$. This reddening is similar to $\textit{E(B$-$V)}=0.077$ reported by \citet{reichardetal03} for the SDSS early DR, but is smaller than $\textit{E(B$-$V)}=0.14$ reported by \citet{gibsonetal09} for the DR5, where in both studies the reddening is derived by comparing the LoBALQ composite to a composite of the whole non-BALQ sample. If the whole non-BALQ sample, rather than the \HeII\ EW matched one, is utilized to calculate the LoBALQ $R$ spectrum, then the \citet{gibsonetal09} result is approximately recovered i.e., $A_{\it V}=0.41$~mag.

\section{Tentative evidence for radiation pressure driving}\label{sec:heii_L_bins}
\begin{figure}
\includegraphics[width=84mm]{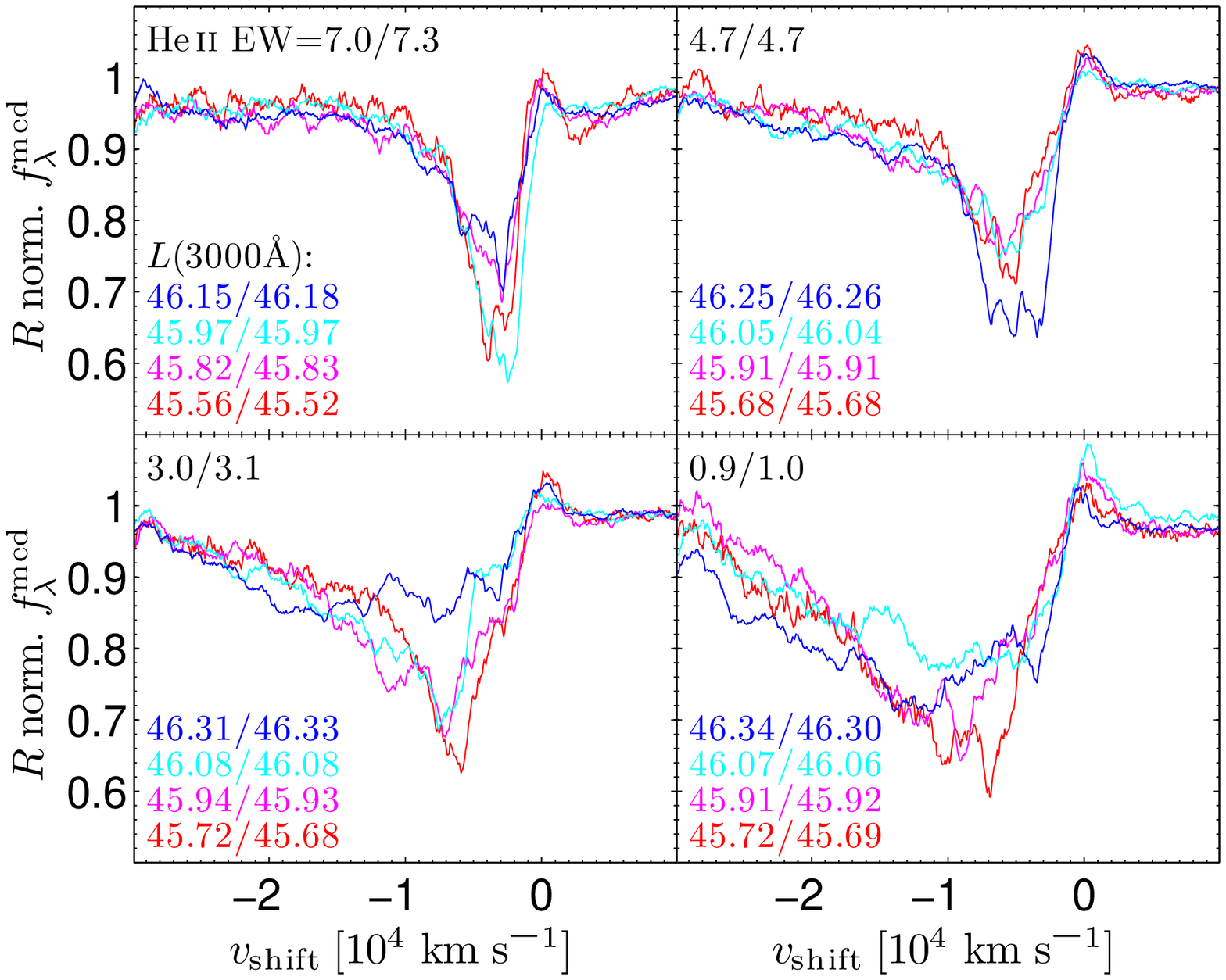}
\caption{The dependence of the \CIV\ BAL profile on the \HeII\ EW and $L(3000\mbox{\AA})$. The BALQ and non-BALQ samples are first binned into \HeII\ EW bins (same as in Fig.~\ref{fig:EW_subs_ab1400}), and then each \HeII\ EW bin is binned based on $L(3000\mbox{\AA})$. The upper left corner in each panel provides the median \HeII\ EW (in \AA) for the BALQ/non-BALQ bins used to generate the plotted ratio. The corresponding values for the four $\log L(3000\mbox{\AA})$ bins (in erg~s$^{-1}$) are indicated in the lower left corner of each panel. Note that for the lower two \HeII\ EW bins (two lower panels), the highest $L(3000\mbox{\AA})$ composite (blue line) has the deepest absorption trough at $\vshift\la-2\times10^4$~\kms.}\label{fig:HeII_L_bins}
\end{figure}

Figure~\ref{fig:HeII_L_bins} presents tentative evidence that the \CIV\ BAL absorber is driven by radiation pressure. The low-$z$ sample is divided into 4$\times$4 bins based on the \HeII\ EW and $L(3000\mbox{\AA})$. The BALQ sample is first divided into four \HeII\ EW `parent' bins, and matched non-BALQ bins are constructed (same procedure as in Fig.~\ref{fig:EW_subs_ab1400}). Then, each \HeII\ EW parent bin is divided into four $L(3000\mbox{\AA})$ bins, with matching non-BALQs from the corresponding \HeII\ EW non-BALQ bin. Each BALQ bin contains by construction equal number of objects ($\sim100$) for all `parent' bins. If the BAL outflow is driven by radiation pressure, then \vshift\ is an increasing function of the force multiplier $\mathcal{M}$ and luminosity $L$, assuming the outflow launching radius is mainly a function of $L$ (e.g., \citealt{laobra02}; see also Section~\ref{sec:phys}). $\mathcal{M}$ is mostly set by the SED, and keeping the \HeII\ EW constant should also keep $\mathcal{M}$ approximately constant i.e., each panel of Fig.~\ref{fig:HeII_L_bins} groups objects with a similar $\mathcal{M}$. The range in $L(3000\mbox{\AA})$ of our sample is too small ($\la1$~dex) to produce highly significant differences in \vshift\ between $L$ bins for a given \HeII\ EW (i.e., $\mathcal{M}$). However, note that for the weakest two \HeII\ EW bins (Fig.~\ref{fig:HeII_L_bins}, two lower panels), the highest $L$ bin reaches larger \vshift\ than the other three bins. This larger \vshift\ for a larger $L$ is expected, if the BAL outflow is driven by radiation pressure (e.g., \citealt*{progaetal98}).

The highest \HeII\ EW bin shows very similar \vshift\ as a function of $L$. Here the peak absorption occurs at $\vshift\sim3000$~\kms, and the absorbing material may still be confined close to the BLR, i.e.\ to the base of the outflow. Overionization occurs just as the outflow starts to build up, before radiation pressure has time to build up a significant outflow component, beyond the initial dispersion in the velocity. Only when the \HeII\ EW is low enough, to allow significant acceleration, the effect of $L$ starts to appear. Clearly, large samples which span a large range in $L$ (e.g.\ \citealt{laobra02}), are needed to explore the radiation pressure interpretation, and clearly separate out the effect of $L$ and the \HeII\ EW on the outflow.

\bsp
\label{lastpage}
\end{document}